# Glutamate regulation of calcium and IP$_3$ oscillating and pulsating dynamics in astrocytes

Maurizio De Pittà[1], Mati Goldberg[1], Vladislav Volman[2,3], Hugues Berry[4], Eshel Ben-Jacob[1,2,*]

1. School of Physics and Astronomy, Tel Aviv University, 69978 Ramat Aviv, Israel

2. Center for Theoretical Biological Physics, UCSD, La Jolla, CA 92093-0319, USA

3. Computational Neurobiology Lab, The Salk Institute, La Jolla, CA 92037, USA

4. Project-Team Alchemy, INRIA Saclay, 91893 Orsay, France

*Corresponding author:

eshel@tamar.tau.ac.il,

Tel.: +972 3 640 7845

Fax: +972 3 642 5787

**Abstract**

Recent years have witnessed an increasing interest in neuron-glia communication. This interest stems from the realization that glia participates in cognitive functions and information processing and is involved in many brain disorders and neurodegenerative diseases. An important process in neuron-glia communications is astrocyte encoding of synaptic information transfer – the modulation of intracellular calcium ($Ca^{2+}$) dynamics in astrocytes in response to synaptic activity. Here, we derive and investigate a concise mathematical model for glutamate-induced astrocytic intracellular $Ca^{2+}$ dynamics that captures the essential biochemical features of the regulatory pathway of inositol 1,4,5-trisphosphate ($IP_3$). Starting from the well-known two-variable (intracellular $Ca^{2+}$ and inactive $IP_3$ receptors) Li-Rinzel model for calcium-induced calcium release, we incorporate the regulation of the $IP_3$ production and phosphorylation. Doing so, we extend it to a three-variable model (which we refer to as the *ChI* model), that could account for $Ca^{2+}$ oscillations with endogenous $IP_3$ metabolism. This *ChI* model is then further extended into the G-*ChI* model to include regulation of the $IP_3$ production by external glutamate signals. Compared with previous similar models, our three-variable models include a more realistic description of the $IP_3$ production and degradation pathways, lumping together their essential nonlinearities within a concise formulation. Using bifurcation analysis and time simulations, we demonstrate the existence of new putative dynamical features. The cross-couplings between $IP_3$ and $Ca^{2+}$ pathways endow the system with self-consistent oscillatory properties and favor mixed frequency-amplitude encoding modes over pure amplitude-modulation ones. These and additional results of our model are in general agreement with available experimental data and may have important implications on the role of astrocytes in the synaptic transfer of information.

# I. Introduction

Astrocytes, the main type of glial cells in the brain, do not generate action potentials like neurons do, yet they can transfer information to other cells and encode information in response to external stimuli by employing "excitable"-like rich calcium ($Ca^{2+}$) dynamics (Volterra and Meldolesi, 2005). Recognition of the potential importance of the intricate inter- and intra-cellular astrocyte dynamics has motivated in recent years, intensive experimental efforts to investigate neuron-glia communication. Consequently, it was discovered that intracellular $Ca^{2+}$ levels in astrocytes can be regulated by synaptic activity (Wang et al., 2006; Pasti et al., 1997; Porter and McCarthy, 1996; Parpura et al., 1994; Dani et al., 1992). Responses to low-intensity synaptic stimulation or spontaneous astrocyte activity usually consist of spatially confined $Ca^{2+}$ transients (Nett et al., 2002; Pasti et al., 1997; Porter and McCarthy, 1996). On the other hand, high-intensity synaptic activity or stimulation of adjacent sites within the same astrocytic process, are generally associated with $Ca^{2+}$ oscillations (Zonta and Carmignoto, 2002) that can bring forth propagation of both intracellular and intercellular waves (Stout et al., 2002; Charles, 1998; Cornell-Bell et al., 1990). Concomitantly, elevation of cytoplasmic $Ca^{2+}$ induces the release from astrocytes of several neurotransmitters (or "gliotransmitters"), including glutamate, ATP or adenosine (see Evanko et al., 2004 for a review). These astrocyte-released gliotransmitters feed back onto pre- and post-synaptic terminals. It implies that astrocytes regulate synaptic information transfer (Volman et al., 2007; Fellin et al., 2004; Araque et al., 1998). Astrocytes can also mediate between neuronal activity and blood circulation (Fellin and Carmignoto, 2004), thus extending neuron-astrocyte communications to the level of neuronal metabolism (Bernardinelli et al., 2004).

The physiological meaning of astrocytic $Ca^{2+}$ signaling remains currently unclear, and a long-standing question is how it participates in the encoding of synaptic information transfer (De Pittà et al., 2008a; 2008b; Volterra and Meldolesi, 2005). Some of the available experimental data suggest a preferential FM (frequency modulation) mode of encoding, namely synaptic activity would be encoded in the frequency of astrocytic $Ca^{2+}$ pulsations (Parpura, 2004). Indeed, cytoplasmic $Ca^{2+}$ waves in astrocytes often appear as pulse-like propagating waveforms (namely pulses of width much smaller than their wavelength), whose frequency increases when the frequency or the intensity of synaptic stimulation grows (Pasti et al., 1997).

Notwithstanding, the possibility of AM (amplitude modulation) encoding of synaptic activity or even of mixed AFM (amplitude and frequency modulation) encoding has also consistently been inferred (Carmignoto, 2000). For instance, the amplitude of $Ca^{2+}$ oscillations in response to external stimuli can be highly variable, depending on the intensity of stimulation (Wang et al., 2006; Finkbeiner, 1993; Cornell-Bell et al., 1990). Experimental evidence suggests that $Ca^{2+}$ dynamics does not simply mirror synaptic activity but is actually much more complex, to a point that astrocytes are suspected of genuine synaptic information processing (Perea and Araque, 2005). The emerging picture is that the properties of $Ca^{2+}$ oscillations triggered by neuronal inputs in astrocytes (including their amplitude, frequency and propagation) are likely to be governed by intrinsic properties of both neuronal inputs and astrocytes (Volterra and Meldolesi, 2005; Pasti et al., 1997).

From the modeling point of view, simplified or two-variable models for intracellular $Ca^{2+}$ signaling can in principle be used to account for the diversity of the observed $Ca^{2+}$ dynamics when the biophysical parameters are varied. We recently presented evidence that one of these two-variable models proposed by Li and Rinzel (Li and Rinzel, 1994) actually predicts that the same cell could encode information about external stimuli by employing different encoding modes. In this model, changes of biophysical parameters of the cell can switch amongst amplitude modulation (AM) of $Ca^{2+}$ oscillations, frequency modulation (FM) of $Ca^{2+}$ pulsations or combined AM and FM (AFM) $Ca^{2+}$ pulsations (De Pittà et al., 2008a; 2008b). We emphasize that one of the cardinal simplifications of the Li-Rinzel model is neglecting the regulation of inositol 1,4,5-trisphosphate ($IP_3$) dynamics, that is its production and degradation. Since $IP_3$ production is regulated by synaptic activity (via extracellular glutamate signaling), $IP_3$ dynamics has to be included for proper modeling of synapse-astrocyte communication. Only such modeling can provide a realistic account of astrocytic $Ca^{2+}$ variations induced by nearby synaptic inputs.

Here, we introduce and investigate a concise model for glutamate-induced intracellular astrocytic dynamics. Using this model we show new putative features of $Ca^{2+}$ dynamics that can have important implications on the role of astrocytes in synapse information transfer. Our model incorporates current biological knowledge related to the signaling pathways leading from extracellular glutamate to intracellular $Ca^{2+}$, via $IP_3$ regulation and $IP_3$-dependent $Ca^{2+}$-induced $Ca^{2+}$ release (CICR). First we extend the Li-Rinzel model to incorporate the regulation of $IP_3$. This yields a three-variable model, called hereafter the "*ChI*" model, for its state variables that are the intracellular $Ca^{2+}$ level *C*, the fraction of inactive $IP_3$ receptors *h*, and the available $IP_3$ concentration *I*. Similar three-variable models have already been introduced in previous works (Kazantsev, 2009; Politi et al., 2006; Höfer et al., 2002; Sneyd et al., 1995; Dupont and Goldbeter, 1993; Meyer and Stryer; 1988; see Falcke, 2004 for a review), yet our modeling includes a more realistic description of $IP_3$ dynamics, in particular with regard to the complex regulatory pathways of $IP_3$ formation and degradation. Furthermore, while we reduce these complex regulatory pathways to a concise mathematical description, we make sure to keep their essential nonlinearities. We then model the contribution of glutamate signals to $IP_3$ production and include this contribution as an additional production term into the $IP_3$ equation of the *ChI* model. We refer to this case as the "G-*ChI*" model.

We utilize bifurcation theory to study the coexistence of various encoding modes of synaptic activity by astrocytes: amplitude modulation (AM), pulsation frequency modulation (FM) and mixed amplitude and frequency modulations (AFM). We also present results of time simulations of the model illustrating the richness of intracellular $Ca^{2+}$ dynamics (hence, of the encoding modes) in response to complex time-dependent glutamate signals.

We note that although the model presented here is derived for the specific case of astrocytes, our approach can be readily adopted to model $Ca^{2+}$ dynamics in other cell types whose coordinated activity is based on intra- and inter-cellular $Ca^{2+}$ signaling, such as heart cells, pancreas cells and liver cells.

## II. Derivation of the three-variable *ChI* model of intracellular $Ca^{2+}$ dynamics

In this section, we describe the concise *ChI* model for intracellular $Ca^{2+}$ dynamics in astrocytes with realistic $IP_3$ regulation. Given the relative intricacy of this signaling pathway (see Figure 1), each basic building block of the model is described separately in the following sections.

**II-1. CICR core and the two-variable Li-Rinzel model**

Intracellular $Ca^{2+}$ levels in astrocytes (as in most other cell types) can be modulated by several mechanisms. These include $Ca^{2+}$ influx from the extracellular space or controlled release from intracellular $Ca^{2+}$ stores such as the endoplasmic reticulum (ER) and mitochondria (Berridge et al., 2000). In astrocytes, though, $IP_3$-dependent calcium-induced calcium release (CICR) from the ER is considered the primary mechanism responsible of intracellular $Ca^{2+}$ dynamics (Agulhon et al., 2008).

Calcium-induced $Ca^{2+}$ release (see Figure 1c) is essentially controlled by the interplay of two specific transports: efflux from the ER to the cytoplasm that is mediated by $Ca^{2+}$-dependent opening of the $IP_3$ receptor ($IP_3R$) channels, and influx into the ER which is due to the action of (Sarco-) Endoplasmic-Reticulum $Ca^{2+}$-ATPase (SERCA) pumps. In basal conditions however (when CICR is negligible), intracellular $Ca^{2+}$ levels are set by the respective contributions of a passive $Ca^{2+}$ leak from the ER, SERCA uptake and plasma membrane $Ca^{2+}$ transport (Li et al., 1994; De Young and Keizer, 1992).

When synaptic activity is large enough, synaptically-released glutamate may spill over the synaptic cleft and bind to the extracellular part of astrocytic metabotropic glutamate receptors (mGluRs) (Porter and McCarthy, 1996). Binding of glutamate to mGluRs increases cytosolic $IP_3$ concentration and promotes the opening of few $IP_3R$ channels (Berridge, 1993). As a consequence, intracellular $Ca^{2+}$ slightly increases. Since the opening probability of $IP_3R$ channels nonlinearly increases with $Ca^{2+}$ concentration (Bezprozvanny, 1991), such initial $Ca^{2+}$ surge increases the opening probability of neighboring channels. In turn this leads to a further increase of cytoplasmic $Ca^{2+}$. These elements therefore provide a self-amplifying release mechanism (hence the denomination of CICR). The autocatalytic action of $Ca^{2+}$ release however reverses at high cytoplasmic $Ca^{2+}$ concentrations, when inactivation of $IP_3R$ channels takes place leading to CICR termination (Iino, 1990). In parallel, SERCA pumps, of which activity increases with cytoplasmic $Ca^{2+}$ (Lytton et al., 1992), quickly sequester exuberant cytoplasmic $Ca^{2+}$ by pumping it back into the ER lumen. The intracellular $Ca^{2+}$ concentration consequently recovers towards basal values which suppress IP$_3$R channels inactivation. Hence, if glutamatergic stimulation is prolonged, intracellular IP$_3$ remains high enough to repeat the cycle and oscillations are observed (Keizer et al., 1995).

The SERCA pump rate can be taken as instantaneous function of cytoplasmic [Ca$^{2+}$] (denoted hereafter by *C*) by assuming a Hill rate expression with exponent 2 (see Appendix A):

$$J_{pump}(C) = v_{ER} \text{Hill}(C^2, K_{ER}) \tag{1}$$

where $v_{ER}$ is the maximal rate of Ca$^{2+}$ uptake by the pump and $K_{ER}$ is the SERCA Ca$^{2+}$ affinity, that is the Ca$^{2+}$ concentration at which the pump operates at half of its maximal capacity (Carafoli, 2002).

The nonspecific Ca$^{2+}$ leak current is assumed to be proportional to the Ca$^{2+}$ gradient across the ER membrane by $r_L$, the maximal rate of Ca$^{2+}$ leakage from the ER:

$$J_{leak}(C) = r_L (C_{ER} - C) \tag{2}$$

where $C_{ER}$ is the Ca$^{2+}$ concentration inside the ER stores (De Young and Keizer, 1992).

IP$_3$R channels can be thought of as ensembles of four independent subunits with three binding sites each: one for IP$_3$ and two for Ca$^{2+}$. The latter include an activation site and a separate site for inactivation (De Young and Keizer, 1992). IP$_3$-binding sensitizes the receptor towards activation by Ca$^{2+}$ but only if both IP$_3$ and activating Ca$^{2+}$ are bound to a fixed set of three out of four subunits, the channel is open.

Assuming that the kinetic rates of the binding reactions are ordered such as IP$_3$-binding>> Ca$^{2+}$-activation>> Ca$^{2+}$-inactivation, Li and Rinzel proposed the following equation for the Ca$^{2+}$ current through the IP$_3$R channels (Li and Rinzel, 1994):

$$J_{chan}(C, h, I) = r_C p^{open} (C_{ER} - C) \tag{3}$$

with the channel open probability that is given by $p^{open} = m_\infty^3 n_\infty^3 h^3$, where $m_\infty = \text{Hill}(I, d_1)$, $n_\infty = \text{Hill}(C, d_5)$ and $h$ account for the three gating reactions, respectively IP$_3$-binding, activating Ca$^{2+}$-binding and Ca$^{2+}$-dependent inactivation of the receptor. The power of 3 was directly suggested by experimental data (De Young and Keizer, 1992; Bezprozvanny et al., 1991). Finally, *I* stands for the intracellular IP$_3$ concentration and $r_C$ is the maximum channel permeability.

Since $Ca^{2+}$ fluxes across the plasma membrane have been proven not necessary for the onset of CICR (Li et al., 1994; Foskett et al., 1991; Rooney et al., 1991), they can be neglected so that the cell-averaged total free $Ca^{2+}$ concentration ($C_0$) is conserved. Hence the ER $Ca^{2+}$ concentration ($C_{ER}$) can be rewritten in terms of equivalent cell parameters as $C_{ER} = \left(\dfrac{C_0 - C}{c_1}\right)$ where $c_1$ is the ratio between the ER and the cytosol volumes. It follows that $J_{chan}$ and $J_{leak}$ can entirely be expressed as functions of cell parameters, namely:

$$J_{chan} = r_C m_\infty^3 n_\infty^3 h^3 (C_0 - (1+c_1)C)$$
$$J_{leak} = r_L (C_0 - (1+c_1)C) \quad (4)$$

Adding together the above terms (equations 1,4), the cytoplasmic $Ca^{2+}$ balance is given by:

$$\dot{C} = (r_C m_\infty^3 n_\infty^3 h^3 + r_L)(C_0 - (1+c_1)C) - v_{ER}\dfrac{C^2}{C^2 + K_{ER}^2} \quad (5)$$

This equation is coupled with an equation for $h$ that accounts for the kinetics of IP$_3$Rs (Li and Rinzel, 1994):

$$\dot{h} = \dfrac{h_\infty - h}{\tau_h} \quad (6)$$

where:

$h_\infty = \dfrac{Q_2}{Q_2 + C}$, $\tau_h = \dfrac{1}{a_2(Q_2 + C)}$, and $Q_2 = d_2 \dfrac{I + d_1}{I + d_3}$.

Equations (5) and (6) form the so-called Li-Rinzel (L-R) model of CICR and constitute the core mechanism of our model for astrocyte $Ca^{2+}$ signaling. We discuss below some of its properties.

**II-2. AM, FM and AFM encoding modes in the Li-Rinzel model**

Calcium acts as a second messenger and transmits information from the extracellular side of the plasma membrane to targets within the cell (Berridge and Bootman, 1997; Jaffe, 1993; Berridge, 1990). In the case of $Ca^{2+}$ signaling in astrocytes however, the information usually arrives as a non-oscillatory stimulus at the plasma membrane and is translated into intracellular $Ca^{2+}$ oscillations. For instance, glutamate concentration at the extracellular side of astrocyte membrane determines the degree of activation of mGluRs and therefore can be directly linked to intracellular IP$_3$ concentration (Verkhratsky and Kettenmann, 1996). It follows that in the L-R model, the level of IP$_3$ can be thought as being directly controlled by glutamate signals impinging on the cell from its external environment. In turn, the level of IP$_3$ determines the dynamics of intracellular Ca$^{2+}$. In physiological conditions glutamate-induced astrocyte Ca$^{2+}$ signaling is synaptically evoked (Wang et al., 2006; Pasti et al., 1997; Porter and McCarthy, 1996). One can therefore think of the Ca$^{2+}$ signal as being an encoding of information about the level of synaptically-released glutamate and ultimately, of synaptic activity. Notably, this information encoding can use amplitude modulation (AM), frequency modulation (FM) or both modulations (AFM) of Ca$^{2+}$ oscillations and pulsations.

We have recently shown that these encoding modes may actually depend on inherent cellular properties (De Pittà et al., 2008a; 2008b). In particular the stronger the SERCA uptake with respect to Ca$^{2+}$ efflux from the ER, the more pulsating and FM-like the encoding. A fast uptake by SERCAs in fact firmly counteracts CICR so that higher Ca$^{2+}$ levels are required for the onset of this latter. When this happens though, the effects of CICR are large and the increase of intracellular Ca$^{2+}$ is fast and remarkable. Accordingly, inactivation of IP$_3$R channels is also faster and basal Ca$^{2+}$ levels are recovered rapidly. In these conditions, the IP$_3$ level modulates the onset of CICR (through $m_\infty$) thus setting the frequency of pulsation (FM encoding). On the contrary the AM case is observed with weaker Ca$^{2+}$ uptakes by SERCAs. Weaker SERCA rates in fact allow for smoother oscillations whose amplitude is mainly dependent on the interplay between CICR onset and Ca$^{2+}$-dependent inactivation. Hence the amplitude of oscillations in these latter conditions depends on IP$_3$ whereas their frequency does not, as it is essentially fixed by IP$_3$R channel recovery from Ca$^{2+}$-dependent inactivation (De Pittà et al., 2008b).

From a dynamical system perspective, AM and FM encoding are associated with well distinct bifurcation diagrams. Amplitude modulations of Ca$^{2+}$ oscillations are typically found when the system exhibits Hopf bifurcations only. In particular when oscillations are born through a supercritical Hopf bifurcation at low IP$_3$ concentration then AM encoding exists (Figures 2a-c). Alternatively if such Hopf bifurcation is subcritical, AFM might be found (De Pittà et al., 2008a). On the contrary, in FM (Figures 2d-f) the presence of a saddle-node homoclinic bifurcation accounts for pulsatile oscillations which rise at arbitrarily small frequency but of amplitude that is essentially independent of the IP$_3$ value (De Pittà et al., 2008b).

Finally, it is important to note that the L-R model assumes that IP$_3$ does not vary with time nor depends on the other variables (that is, its concentration $I$, in equations (5) and (6), is a parameter of the model). Yet examination of the underlying signaling pathways (Figure 1) immediately hints that IP$_3$ concentration indeed depends on both intracellular Ca$^{2+}$ and extracellular glutamate, so that IP$_3$ should be an additional variable in the model. Our aim in the present article is to devise a model that incorporates these dependencies.

**II-3. IP$_3$ regulation: the *ChI* model**

*a) PLC$\delta$ production*

In astrocytes, IP$_3$ together with diacylglycerol (DAG) is produced by hydrolysis of phosphatidylinositol 4,5-bisphosphate (PIP$_2$) by two phosphoinositide-specific phospholipase C (PLC) isoenzymes, PLC$\beta$ and PLC$\delta$ (Rebecchi and Pentyala, 2000). The activation properties of these two isoenzymes are different and so likely are their roles. PLC$\beta$ is primarily controlled by cell surface receptors, hence its activity is linked to the level of external stimulation (i.e. the extracellular glutamate) and as such, it pertains to the glutamate-dependent IP$_3$ metabolism and will be addressed in the next section.

On the contrary PLC$\delta$ is essentially activated by increased intracellular Ca$^{2+}$ levels (Figure 1d) (Rhee and Bae, 1997). Structural and mutational studies of complexes of PLC$\delta$ with Ca$^{2+}$ and IP$_3$ revealed complex interactions of Ca$^{2+}$ with several negatively charged residues within its catalytic domain (Rhee, 2001; Essen et al., 1997; Essen et al., 1996;), a hint of cooperative binding of Ca$^{2+}$ to this enzyme. In agreement with these experimental findings, the PLC$\delta$ activation rate can be written as (Höfer et al., 2002; Pawelczyk and Matcki, 1997):

$$v_\delta(C, I) = v'_\delta(I) \cdot \mathrm{Hill}(C^2, K_{PLC\delta}) \qquad (7)$$

where the maximal rate of activation depends on the level of intracellular IP$_3$. Experimental observations shows that high (>1 µM) IP$_3$ concentrations inhibit PLCδ activity by competing with PIP$_2$ binding to the enzyme (Allen et al., 1997). Accordingly, assuming competitive binding (Stryer, 1999), the maximal PLCδ-dependent IP$_3$ production rate can be modeled as follows:

$$v'_\delta(I) = \frac{\bar{v}_\delta}{1 + \frac{I}{\kappa_\delta}} \qquad (8)$$

where $\kappa_\delta$ is the inhibition constant of PLCδ activity.

Figure 3 shows the behavior of this term when Ca$^{2+}$ and corresponding IP$_3$ levels obtained from the bifurcation diagrams of the L-R model in Figure 2 are substituted into equation (7). We have set $K_{PLC\delta}$ to a value that is close to the Ca$^{2+}$ concentration of the lower bifurcation point. This allows us to translate the large-amplitude Ca$^{2+}$ oscillations into oscillations of $v_\delta$ that could preserve the main AM/FM properties.

*b) IP$_3$ degradation*

Two major IP$_3$ degradation pathways have been described so far (Figure 1b). The first one is through dephosphorylation of IP$_3$ by inositol polyphosphate 5-phosphatase (IP-5P). The other one occurs through phosphorylation of IP$_3$ by the IP$_3$ 3-kinase (IP$_3$-3K) and is Ca$^{2+}$ dependent (Zhang et al., 1993).

The rate of both IP-5P dephosphorylation ($v_{5P}$) and IP$_3$-3K phosphorylation ($v_{3K}$) of IP$_3$ can be considered as of Michaelis-Menten type (Dupont and Erneux, 1997; Togashi et al., 1997; Irvine et al., 1986). Therefore:

$$\begin{aligned} v_{5P}(I) &= \bar{v}_{5P} \cdot \text{Hill}(I, K_5) \\ v_{3K}(C, I) &= v^*_{3K}(C) \cdot \text{Hill}(I, K_3) \end{aligned} \qquad (9)$$

Since $K_5$>10 µM (Sims et Allbritton, 1998; Verjans et al., 1992;), and physiological levels of IP$_3$ are in general below this value, IP-5P is likely not to be saturated by IP$_3$. It follows that the rate of IP$_3$ degradation by IP-5P can be linearly approximated:

$$v_{5P}(I) \approx \bar{r}_{5P} \cdot I \qquad (10)$$

where $\bar{r}_{5P}$ is the linear rate of IP$_3$ degradation by IP-5P and can be defined by parameters in equation (9) as $\bar{r}_{5P} = \bar{v}_{5P}/K_5$.

In basal conditions, phosphorylation of IP$_3$ by IP$_3$-3K is very slow. The activity of IP$_3$-3K is substantially stimulated by Ca$^{2+}$/calmodulin (CaM) via CaMKII-catalyzed phosphorylation (Figure 1b) (Communi et al., 1997). However, other experimental reports have suggested that Ca$^{2+}$-dependent PKC phosphorylation of IP$_3$-3K could have inhibitory effects (Sim et al., 1990). Notwithstanding, evidences for this latter possibility are contradictory (Communi et al., 1995). Hence, for the sake of simplicity, we have chosen in the present model to consider the simplified case were only CaMKII-catalyzed phosphorylation of IP$_3$-3K is present (Figure 1f).

Phosphorylation of IP$_3$-3K by active CaMKII (CaMKII*) only occurs at a single threonine residue (Communi et al., 1999; Communi et al., 1997), therefore we can assume that $v_{3K}^*(C) \propto [\text{CaMKII*}]$. Activation of CaMKII is Ca$^{2+}$/CaM-dependent and occurs in a complex fashion because of the unique structure of this kinase, which is composed of ~12 subunits with three to four phosphorylation sites each (Kolodziej et al., 2000). Briefly, Ca$^{2+}$ elevation leads to the formation of a Ca$^{2+}$-CaM complex (CaM$^+$) that may induce phosphorylation of some of the sites of each CaMKII subunit. CaMKII quickly and fully activates when two of these sites (at proximal subunits) are phosphorylated (Hanson et al, 1994). In spite of the occurrence of multiple CaM$^+$ binding to the inactive kinase, experimental investigations showed that KII activation by CaM$^+$ can be approximated by a Hill equation with unitary coefficient (De Konick and Schulman, 1998). Hence, if we surmise the following kinetic reaction scheme for CaMKII phosphorylation:

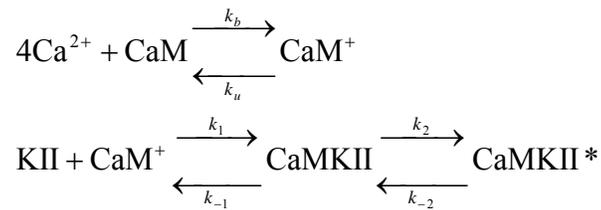

it can be demonstrated that $[\text{CaMKII*}] \propto \text{Hill}(C^4, K_D)$ (see Appendix B).

Accordingly, $v_{3K}^*(C) \propto \text{Hill}(C^4, K_D)$ and the equation for IP$_3$-3K-dependent IP$_3$ degradation reads:

$$v_{3K}(C,I) = \bar{v}_{3K} \cdot \text{Hill}(C^4, K_D) \cdot \text{Hill}(I, K_3) \tag{12}$$

Experimental observations show the existence of three regimes of IP$_3$ metabolism (Sims and Allbritton, 1998). At low [Ca$^{2+}$] and [IP$_3$] (<400 nM and <1 µM respectively), IP-5P and IP$_3$-3K degrade roughly the same amounts of IP$_3$. Then at high [Ca$^{2+}$] (≥400 nM) but low [IP$_3$] (≤8 µM), IP$_3$ is predominantly metabolized by IP$_3$-3K. Eventually for [IP$_3$] greater than 8 µM, when IP$_3$-3K activity saturates, IP-5P becomes the dominant metabolic enzyme, independently of [Ca$^{2+}$].

In our modeling, the third regime – corresponding to [IP$_3$]>8 µM – exceeds the range of validity for the linear approximation of IP-5P degradation (equation 10) and therefore cannot be taken into account. However, it can be shown that the first two regimes are sufficient to reproduce Ca$^{2+}$ oscillations and pulsations, thus restricting the core features of IP$_3$ metabolism to the maximal rates of IP$_3$ degradation by IP$_3$-3K and IP-5P and to the Ca$^{2+}$ dependence of IP$_3$-3K. In particular, by opportune choice of parameters such as $\bar{v}_{3K} > K_3 \bar{r}_{5P}$, theoretical investigations showed that these two regimes are essentially brought forth by the Ca$^{2+}$-dependent Hill term in the expression of $v_{3K}$, irrespectively of the assumption of Michaelis-Menten kinetics for IP$_3$-dependence of IP$_3$-3K (Figure 4). Accordingly, a linear approximation for $v_{3K}$ such as:

$$v_{3K}(C,I) = \bar{r}_{3K} \cdot \text{Hill}(C^4, K_D) \cdot I \tag{13}$$

where $\bar{r}_{3K} = \bar{v}_{3K}/K_3$, could also be considered instead of equation (12), in agreement with previous investigations found in literature (Politi et al., 2006; Sims et Allbritton, 1998).

Indeed, the behaviors of $v_{3K}$ in equations (12) and (13) for IP$_3$ and Ca$^{2+}$ concentrations obtained from the corresponding Li-Rinzel bifurcation diagrams are qualitatively similar (Figure 5). Moreover, the overall bifurcation diagrams are largely conserved (results not shown). The main quantitative difference is that the linear approximation yields stronger degradation rates. In particular, the IP$_3$-3K rate can be up to 2-fold higher in equation (13) than in equation (12). This is particularly marked when high [Ca$^{2+}$] are reached, such as in FM conditions (Figures 5c-d). Notwithstanding, the Michaelis-Menten constant of IP$_3$-3K for its substrate is experimentally reported to be $K_3 \approx 1$ µM (Takazawa et al., 1989; Irvine et al., 1986) and it is likely that intracellular IP$_3$ levels can reach such micromolar concentrations *in vivo* (Mishra and Bhalla, 2002). Therefore, in the following, we will keep the Michaelis-Menten formulation for $v_{3K}$ (equation 12).

Finally, experimental measurements show that for [Ca$^{2+}$]>1 µM and low IP$_3$ levels, the IP$_3$-3K activity exceeds that of IP-5P by almost 20-fold. In the model, this means that if $I \ll K_3$ (i.e. $v_{3K}(C,I) \approx \bar{r}_{3K} \cdot I$), then $v_{3K} \approx 20 v_{5P}$. Accordingly, we set the maximal degradation rates in the following such that $\bar{v}_{3K} \approx 20 K_3 \bar{r}_{5P}$.

*c) Model analysis*

In summary, our model of Ca$^{2+}$ dynamics with endogenous IP$_3$ metabolism is based on the two L-R equations (equations 5-6), but the IP$_3$ concentration (*I*) is now provided by a third coupled differential equation (summing the terms given by equations 7,10,12):

$$\dot{I} = \frac{\bar{v}_\delta}{1+\dfrac{I}{\kappa_\delta}} \mathrm{Hill}(C^2, K_{PLC\delta}) - v_{3K}\, \mathrm{Hill}(C^4, K_D)\, \mathrm{Hill}(I, K_3) - r_{5P} I \quad (14)$$

Equation (14) together with equations (5) and (6) define our three-variable "*ChI*" model, in the name of the letters used to denote its state variables.

Consistency of the *ChI* model with respect to the L-R core model was sought by comparing two curves for pseudo-steady states. First, we set $\dot{I} = 0$ and $C \to 0$ in equation (14), and solved for *I* as a function of *C* in the resulting equation. In parallel, we set $\dot{C} = 0$ in equation (5) and solved for *I* as a function of *C* in the resulting equation as well. The two resulting *I-C* curves should be as similar as possible. Analysis showed they are indeed relatively similar (Figure 6) if one chooses $K_{PLC\delta} \leq H_1$, $K_D \approx H_2$, $K_3 > H_2$, where $H_1$ and $H_2$ denote Ca$^{2+}$ and IP$_3$ concentrations at the two Hopf bifurcations in the L-R bifurcation diagrams (Figure 2). Such choice of parameters together with the others given in Table 1 ensures the existence of Ca$^{2+}$ and IP$_3$ oscillations with amplitudes that are in agreement with those reported in literature (Mishra and Bhalla, 2002) (see Figure 3 of Online Supplementary Material).

An important feature of our model is that despite the coupling between Ca$^{2+}$ and IP$_3$, the equation for Ca$^{2+}$ dynamics (equation 5) does not contain parameters found within the equation of IP$_3$ dynamics (equation 14). This means that the equation of the *C*-nullcline does not change with respect to the L-R model. Because the shape of this nullcline is crucial for the encoding mode (see Figures 2a,c), the occurrence of AM, FM or AFM modes in the *ChI* model is essentially established by the parameters of the L-R core model.

The only possible way that IP$_3$ metabolism could affect the encoding mode is by modulating the dynamics of the channel inactivation variable *h*. This mechanism is suggested by the projection of the surfaces for $\dot{C}=0$, $\dot{h}=0$ and $\dot{I}=0$ (Figure 7) onto the *C-I* plane for different values of *h* and *C* (Figure 8). We note indeed that the *C*-nullcline depends on the value of *h* but not the *I*-nullcline. On the contrary, both *h*-nullcline and *I*-nullcline change with *C,* which hints that the coupling between Ca$^{2+}$ and IP$_3$ dynamics essentially occurs through *h*. We may expect that since *h* sets the slow time scale of the oscillations, the effect of IP$_3$ metabolism on Ca$^{2+}$ dynamics in our model is mainly a modulation of the oscillation frequency. This aspect is further discussed in Sections IV and V, following the introduction in the next section of the last term of our model, namely the glutamate-dependent IP$_3$ production.

## III. Modelling glutamate regulation of IP$_3$ production: the G-*ChI* model

The contribution of glutamate signals to IP$_3$ production can be taken into account as an additional production term in the IP$_3$ equation of the above three-variable *ChI* model. The resulting new model is referred to as the "G-*ChI*" model.

Glutamate-triggered Ca$^{2+}$ signals in astrocytes are mediated by group I and II mGluRs (Zur Niedem and Dietmer, 2006). Metabotropic GluRs are G-protein coupled receptors associated with the phosphotidylinositol signaling-cascade pathway (Teichberg, 1991). Although it is likely that the type of mGluRs expressed by astrocytes depends on the brain region and the stage of development (Gallo and Ghiani, 2000), it seems reasonable to assume that such differences are negligible in terms of the associated second-messenger pathways (Abe et al., 1992; Masu et al., 1991).

The G protein associated with astrocyte mGluRs is a heterotrimer constituted by three subunits: *α, β* and *γ.* Glutamate binding to mGluR triggers receptor-catalyzed exchange of GTP from the G*βγ* subunits to the G*α* subunit. The GTP-loaded G*α* subunit then dissociates from the G protein in the membrane plane and binds to a co-localized PLCβ (Figures 1a,e). Upon binding to Gα, the activity of PLCβ substantially increases, thus promoting PIP$_2$ hydrolysis and IP$_3$ production. Activation of PLCβ can therefore, at first approximation, be directly linked to the number of bound mGluRs, hence to the level of external stimulation. It follows that glutamate-dependent IP$_3$ production can be written in the following generic form:

$$v_{glu}(\gamma,C) = \bar{v}_\beta \cdot R(\gamma,C) \qquad (15)$$

where $\bar{v}_\beta$ is the maximal PLCβ rate, that depends on the surface density of mGluRs, and $R(\gamma,C)$ is the fraction of activated (bound) mGluRs. Experimental evidence shows that PLCβ activity (*i.e.* $\bar{v}_\beta$ in equation 15) is also dependent on intracellular Ca$^{2+}$ (Rhee and Bae, 1997). Notwithstanding, such dependence seems to occur for [Ca$^{2+}$]>10 µM, hence out of our physiological range (Allen et al., 1997). Therefore, Ca$^{2+}$ dependence of PLCβ maximal rate will not be considered here.

$R(\gamma,C)$ can be expressed in terms of extracellular glutamate concentration (γ) at the astrocytic plasma membrane, assuming a Hill binding reaction scheme, with an exponent ranging between 0.5 and 1 (Suzuki et al, 2004). In the current study, we choose 0.7 yielding:

$$R(\gamma,C) = \text{Hill}\left(\gamma^{0.7}, K_\gamma(\gamma,C)\right) \qquad (16)$$

In equation (16), $R(\gamma,C)$ is expressed as a Hill function with a midpoint that depends on glutamate and intracellular Ca$^{2+}$ concentrations. This choice was motivated by the termination mechanism of PLCβ signaling, that occurs essentially through two reaction pathways (Rebecchi and Pentyala, 2000): (*i*) reconstitution of the inactive G-protein heterotrimer due to the intrinsic GTPase activity of activated Gα subunits and (*ii*) PKC phosphorylation of the receptor, or of the G protein, or of PLCβ or some combination thereof. We lump both effects into a single term, $K_\gamma(\gamma,C)$, such as the effective Hill midpoint of $R(\gamma,C)$ increases as PLCβ termination takes over, namely:

$$K_\gamma(\gamma,C) = K_R\left(1 + \frac{K_p}{K_R}\text{Hill}(\gamma^{0.7},K_R)\text{Hill}(C,K_\pi)\right) \qquad (17)$$

Here $K_R$ is the Hill midpoint of glutamate binding with its receptor whereas $K_p$ measures the increment of the apparent affinity of the receptor due to PLCβ terminating signals.

Hill($\gamma^{0.7}$, $K_R$) accounts for the intrinsic GTPase-dependent PLCβ activity termination, as this effect is linked to the fraction of activated Gα subunits and therefore can be put in direct proportionality with the fraction of bound receptors.

Hill($C$, $K_\pi$) instead accounts for PKC-related phosphorylation-dependent termination of PLCβ activity. Experimental data suggest that the target of PKC in this case is either the G protein or PLCβ itself (Ryu et al., 1990). Generally speaking, phosphorylation by PKC may modulate the efficiency of ligand-binding by the receptor, the coupling of occupied receptor to the G protein, or the coupling of the activated G protein to PLCβ (Fisher, 1995). All these effects indeed are lumped into equation (17), as explained below.

PKC is activated in a complex fashion (Figure 1e). Indeed its activation by mere intracellular $Ca^{2+}$ is minimal (Nishizuka, 1995), while full activation is obtained by binding of the coactivator DAG. In agreement with this description, PKC activation can be approximated by a generic Hill reaction scheme whereas $Ca^{2+}$ dependent PKC phosphorylation can be assumed Michaelis-Menten (Ryu et al., 1990) so that [PKC*] $\propto$ Hill([DAG], $K'_{DAG}$)·Hill($C$, $K_\pi$). Remarkably, [DAG] can itself be related to intracellular $Ca^{2+}$ concentration (Codazzi et al., 2001) so that [PKC*] can be rewritten as [PKC*] $\propto$ Hill($C$, $K_{DAG}$)·Hill($C$, $K_\pi$). Finally, $K_{DAG} \ll K_\pi$ (Codazzi et al., 2001; Nishizuka, 1995; Shinomura et al., 1991) so that we can eventually approximate the product of the two Hill functions by that with the highest midpoint (see Appendix A for the derivation of this approximation). That yields: [PKC*] $\propto$ Hill($C$, $K_\pi$), which accounts for the second Hill function in equation (17).

To complete the model, it can be shown by numerical analysis of $K_\gamma(\gamma, C)$ (equation 17) that the term related to the GTPase-dependent PLCβ termination pathway, i.e. Hill($\gamma^{0.7}$, $K_R$), can be neglected at first approximation (Figure 9). Hence $K_\gamma(\gamma, C)$ can be simplified as $K_\gamma(C)$:

$$K_\gamma(C) \approx K_R \left(1 + \frac{K_p}{K_R} \text{Hill}(C, K_\pi)\right) \quad (18)$$

Using equations (15), (16) and (18), our final expression for the glutamate-dependent $IP_3$ production reads:

$$v_{glu}(\gamma,C) = \bar{v}_\beta \cdot \text{Hill}\left(\gamma^{0.7}, K_R\left(1 + \frac{K_p}{K_R}\text{Hill}(C, K_\pi)\right)\right) \tag{19}$$

Including equation (19) into (14), we obtain

$$\dot{I} = \bar{v}_\beta \cdot \text{Hill}\left(\gamma^{0.7}, K_R\left(1 + \frac{K_p}{K_R}\text{Hill}(C, K_\pi)\right)\right) + \frac{\bar{v}_\delta}{1 + \frac{I}{\kappa_\delta}}\text{Hill}(C^2, K_{PLC\delta}) + \\ - v_{3K}\text{Hill}(C^4, K_D)\text{Hill}(I, K_3) - r_{5P}I \tag{20}$$

This equation combined with equations (5) and (6) define our G-*ChI* model of glutamate-dependent intracellular Ca$^{2+}$ dynamics in astrocytes.

## IV. Dynamical behaviors and coding modes of the G-*ChI* model

The dynamical features of the G-*ChI* model for different extracellular concentrations of glutamate can be appreciated by inspection of bifurcation diagrams in Figures 10 and 11. We note that the choice of $\bar{v}_\beta$, the maximal rate of glutamate-dependent IP$_3$ production which is linked to the density of receptors on the extracellular side of the astrocyte membrane, can substantially influence the bifurcation structure of the model and the extent of the oscillatory range. Indeed, as $\bar{v}_\beta$ decreases the oscillatory range expands towards infinite glutamate concentrations but the amplitude of oscillations concomitantly decreases (at least with regard to the IP$_3$ concentration).

The extension of the oscillatory range is due to the shift towards infinity of the subcritical Hopf bifurcation at high glutamate concentrations (compare Figures 11a,d). Notably, for some values of receptor density there seems to be coexistence of oscillations and asymptotic stability at high concentrations of extracellular glutamate, depending on the state of the cell prior to the onset of stimulation (Figures 11d-f).

As $\bar{v}_\beta$ decreases, degradation becomes progressively preponderant so that IP$_3$ peak levels are lower and the IP$_3$R channels open probability is also reduced. Consequently CICR is weaker and the increase of cytosolic Ca$^{2+}$ is smaller. Then Ca$^{2+}$-dependent PKC activation is reduced and termination of PLCβ signaling by PKC-dependent phosphorylation is limited. Moreover, if saturation of receptors occurs (i.e. $R(\gamma,C) \approx 1$) and oscillations are observed in this case, it follows that higher extracellular glutamate concentrations cannot further affect the intracellular $Ca^{2+}$ dynamics.

The value of $\bar{v}_\beta$ at which intracellular $Ca^{2+}$ dynamics locks onto stable oscillations also depends on $\bar{v}_\delta$, the strength of the endogen PLCδ-mediated $IP_3$ production. To some extent, increasing $\bar{v}_\delta$ decreases the minimal $\bar{v}_\beta$ value above which oscillations appear, provided that CICR is strong enough to activate enough PLCδ to keep $IP_3$ levels above the lower Hopf bifurcation (results not shown).

Coupling between $IP_3$ and $Ca^{2+}$ dynamics in the G-*ChI* model might have important implications for the encoding of the stimulus. Bifurcation diagrams in Figures 10 and 11 were derived using different sets of parameters that pertain respectively to AM and FM encoding in the *ChI* model as well as in the L-R core model (see Table 1 and Figure 3 in Online Supplementary Material). Notwithstanding, the applicability of these definitions to the G-*ChI* model might lead now to some ambiguity.

We have previously assumed that AM (FM) encoding exists only if the amplitude (frequency) of oscillations (pulsations) throughout the oscillatory range can at least double with respect to its minimum value (De Pittà et al., 2008b). Here, if we consider the AM-derived bifurcation diagrams for $Ca^{2+}$ and $IP_3$ dynamics (Figure 10), we note that AM is still found since oscillations rise with arbitrarily small amplitude for the supercritical Hopf point at lower stimulus intensity (Figures 10a-b,d-e). But the period of oscillations (Figures 10c,f) at the upper extreme of the oscillatory range is almost half that observed at the onset of oscillations at the lower Hopf point. Thus FM also occurs. Notably, in such conditions, oscillations resemble pulsating dynamics. In other words, rather than pure AM encoding, as we could expect by a set of parameters that provides AM in the *ChI* model (Figures 1a-c in Online Supplementary Material), it seems that in the G-*ChI* model, $Ca^{2+}$ oscillations become AFM encoding. Notably, $IP_3$ dynamics appears to be always AFM encoding both in the AM (Figures 10b,e) and in the FM-derived bifurcation diagrams (Figures 11b,f,i).

Conversely, mere FM encoding is essentially preserved for $Ca^{2+}$ dynamics derived from FM encoding sets of parameters in the *ChI* model, although a significant increase of the range of amplitudes of pulsations can be pointed out (compare Figure 11d with Figure 3d in Online Supplementary Material). These observations indicate that the *G-ChI* model accounts either for FM or AFM encoding $Ca^{2+}$ oscillations, which are though always coupled with AFM encoding $IP_3$ oscillations. In addition they provide further support to the above-stated notion that $IP_3$ metabolism could consistently modulate the frequency of $Ca^{2+}$ pulsating dynamics more than their amplitude (see Section II-3c).

On the contrary, the amplitude and shape of $IP_3$ oscillations appear to be dramatically correlated to that of $Ca^{2+}$ oscillations, as a consequence of the numerous $Ca^{2+}$-dependent feedbacks on $IP_3$ metabolism. Smooth $Ca^{2+}$ oscillations such as those obtained in AM-like conditions (Figure 12a, AM) are coupled with small zigzag $IP_3$ oscillations (Figure 12b, AM). Under FM conditions instead, pulsating large-amplitude $Ca^{2+}$ variations (Figure 12a, FM) can be lagged by analogous $IP_3$ oscillations (Figure 12b, FM), with the difference that whereas $Ca^{2+}$ pulsations are almost fixed in their amplitude, $IP_3$ ones can substantially vary.

Simulations of physiologically equivalent glutamate stimulation and associated astrocyte $Ca^{2+}$–$IP_3$ patterns are shown in Figure 13. Real MEA-recording data were considered as inputs of a single glutamatergic synapse (modeled as in Tsodyks et al., 1997) and a fraction of the released glutamate was assumed to impinge the astrocyte described by our model. We may notice that from the stimulus up to $Ca^{2+}$ dynamics the smoothness of the patterns seems to increase. Indeed, the highly jagged glutamate stimulus turns into a less indented $IP_3$ signal which is coupled with even smoother $Ca^{2+}$ oscillations. Depending on the inherent cellular properties (Figure 13, for example, considers two cases associated with different SERCA $Ca^{2+}$ affinities), the difference of smoothness between $IP_3$ and $Ca^{2+}$ can be dramatic, more likely in the case of FM encoding $Ca^{2+}$ pulsations (compare Figures 13a-b, AM and FM).

## V. Discussion

Calcium dynamics in astrocytes can be driven by extra-cellular signals (such as glutamate neurotransmitter) through regulation of the intracellular $IP_3$ levels. Therefore, a pre-requisite towards unraveling the response of astrocytes to such signals is a thorough understanding of the complex IP$_3$-related metabolic pathways that regulate intracellular Ca$^{2+}$ dynamics. Here, we have devised and studied a model for agonist-dependent intracellular Ca$^{2+}$ dynamics that captures the essential biochemical features of the complex regulatory pathways involved in glutamate-induced IP$_3$ and Ca$^{2+}$ oscillations and pulsations. Our model is simple, yet it retains the essential features of the underlying physiological processes that constitute the intricate IP$_3$ metabolic network.

More specifically, the equation for IP$_3$ dynamics is a central component of our model because of the large number of metabolic reactions that it accounts for and because coupling with intracellular Ca$^{2+}$ dynamics is resolved through complex feedback mechanisms. Production of IP$_3$ depends on the agonist/receptor-dependent PLCβ activation as well as on the endogenous agonist-independent contribution of PLCδ because both isoenzymes are found in astrocytes (Rebecchi and Pentyala, 2000).

We linked the relative expression of these two isoenzymes to the expression of PKC and to the strength of PLCβ regulation by PKC. Indeed, Ca$^{2+}$-dependent PKC activation can phosphorylate the receptor or PLCβ or a combination thereof, leading to termination of IP$_3$ production (Ryu et al., 1990). In astrocytes, this mechanism has been suggested to limit the duration of Ca$^{2+}$ oscillations thus defining their frequencies (Codazzi et al., 2001). In agreement with this idea, a stronger PKC-dependent inhibition of PLCβ shrank the oscillatory range in our model astrocyte and led to the progressive loss of long-period oscillations.

In our model, the PKC-dependent inhibition of PLCβ is counteracted if PLCδ expression is high enough to support high IP$_3$ production levels and the resulting release of Ca$^{2+}$ from the intracellular stores. This observation raises the possibility of phase-locked Ca$^{2+}$ oscillations under conditions of intense stimulation. Phase-locked Ca$^{2+}$ oscillations were also found in other models of agonist-dependent intracellular Ca$^{2+}$ dynamics (Chay et al., 1995a; Chay et al., 1995b; Cuthbertson and Chay, 1991) and are often associated with pathological conditions (Uhlhaas and Singer, 2006; Shrier et al., 1987). In our model, persistent pulsating Ca$^{2+}$ dynamics that are essentially independent of the level of stimulation are observed for weak maximal rates of IP$_3$ production by PLCβ (Figures 10d-e,11g-h). In astrocytes, such persistent oscillations could also be interpreted as a fingerprint of pathological conditions (and Meldolesi, 2005; Balászi et al., 2003). In fact, a decay of PLCβ activity is likely to occur for instance, if the density of effective metabotropic receptors in the astrocytic plasma membrane decreases, such as in the case of epileptic patients with Ammon's Horn sclerosis (Seifert et al., 2004).

We note that although focusing on stimulus-triggered $Ca^{2+}$ oscillations, our study also hints a possible link between modulation of frequency and amplitude of $Ca^{2+}$ pulsations and spontaneous $Ca^{2+}$ dynamics. Recently it has been shown that the inter-pulse interval of the spontaneous $Ca^{2+}$ oscillations is inherently stochastic (Skupin and Falcke, 2008; Skupin et al., 2008). In particular, experimental observations are compatible with model studies of a local stochastic nucleation mechanism that is amplified by the spatial coupling among $IP_3R$ clusters through $Ca^{2+}$ or $IP_3$ diffusion (Skupin et al., 2008; Falcke, 2003). Our analysis may provide meaningful clues to identify what factors and processes within the cell could affect the rate of wave nucleation. More specifically we may predict that putative intracellular $IP_3$ dynamics could affect the statistics of $Ca^{2+}$ inter-pulse intervals not only in terms of spatial coupling amongst $IP_3R$ clusters by means of intracellular $IP_3$ gradients, but also by modulation of either the recovery from $Ca^{2+}$ inhibition or the progressive sensitization of $IP_3Rs$ by $Ca^{2+}$ (Tang and Othmer, 1995). The resulting scenario therefore, would still be that of a local stochastic nucleation mechanism amplified by $IP_3R$ spatial coupling, but the local $IP_3R$ and SERCA parameters would vary according with to the biochemical regulation system presented in the current work.

A critical question in experiments is the identification of the mechanism that drives $IP_3$ oscillations and pulsations (Young et al., 2003; Nash et al., 2001; Hirose et al., 1999; Sims and Allbritton, 1998). In our model, self-sustained $IP_3$ oscillations are brought about by the coupling of $IP_3$ metabolism with $Ca^{2+}$ dynamics. In other words, our model can be considered as a self-consistent astrocytic generator of $Ca^{2+}$ dynamics. This might have broad implications for astrocyte encoding of information and neuron-glia communication. We previously demonstrated that modulation by astrocytes of synaptic information transfer could account for some of the peculiar dynamics observed in spontaneous activity of cultured cortical networks (Volman et al., 2007). In particular, a simple neuron-glia circuit composed of an autaptic neuron "talking" with a proximal astrocyte could serve as a self-consistent oscillator when fed by weak external signals. The results presented in the current study suggest an alternative, more robust, way (independent of synaptic architecture) to form glia-based self-consistent oscillators. The relative contribution and significance of either the astrocytic or the $IP_3$-based hypotheses to the spontaneous network activity, need to be assessed by future combined experiments and modeling. Meanwhile, the analysis of our present model suggests that in astrocytes, different second messenger molecules are engaged in an intricate dialogue, likely meaning that those non-neural cells might be crucially important to decipher some of the enigmas of neural information processing.

Another significant prediction of our model is that $IP_3$ dynamics is essentially AFM and $Ca^{2+}$ oscillations/pulsations are inherently FM encoding, that is they can be either FM or AFM but not AM (Berridge, 1997; De Pittà et al, 2008a; 2008b). In FM, $Ca^{2+}$ oscillations resemble pulses. In AFM instead, their shape is smoother and necessarily depends on the stimulus dynamics.

The assumption that $IP_3$ oscillations are always AFM-encoding could provide an optimal interface between agonist stimuli and intracellular $Ca^{2+}$ signals. The stimuli impinge the cell in the form of trains of pulses or burst of pulses and information is carried in the timing of these pulses rather than in their amplitude (Sejnowski and Paulsen, 2006). AFM features in $IP_3$ signals could perfectly match these stimuli, embedding the essential features of the spectrum of the signal into the spectrum of the $IP_3$ transduction. Hence, $IP_3$ signaling with FM features could offer an efficient way to keep the essence of the information of the stimulus. On the other hand, because $Ca^{2+}$ signals are triggered primarily by sufficiently ample elevations of $IP_3$ (Li et al., 1994), the coexistence of AM features within the $IP_3$ signal seems to be a necessary prerequisite in order to trigger CICR.

The fact that coupling of $IP_3$ metabolism with CICR does not allow pure AM-encoding is in general agreement with experimental data on intracellular $Ca^{2+}$ signaling in several cells (Berridge, 1998; Woods et al., 1986) including astrocytes (Pasti et al., 1997).

Notwithstanding the possibility of AFM encoding $Ca^{2+}$ oscillations has recently come up as a reliable alternative mechanism to explain gliotransmitter exocytosis that is dependent on the specific agonist that triggers astrocyte $Ca^{2+}$ dynamics (Montana et al., 2006; Carmignoto, 2000).

The above could be relevant to understand the origin of the integrative properties of $Ca^{2+}$ signaling in astrocytes (Perea and Araque, 2005a). Our analysis in fact shows that such properties could result from at least two steps of integration, one at the transduction of the agonist signal into $IP_3$ signal and the other at the cross-coupling between $IP_3$ and $Ca^{2+}$ signals. Indeed AFM-encoding $IP_3$ dynamics could deploy smoothing of the highly indented agonist stimulus thus hinting possible integrative properties for $IP_3$ signals (Figure 13). On the other hand, the associated $Ca^{2+}$ patterns look even smoother suggesting a further integration step that likely relies only on the inherent features of CICR.

## Acknowledgements

The authors wish to thank Vladimir Parpura, Giorgio Carmignoto and Ilyia Bezprozvanny for insightful conversations. V. V. acknowledges the support of the U.S. National Science Foundation I2CAM International Materials Institute Award, Grant DMR-0645461. This research was supported by the Tauber Family Foundation, by the Maguy-Glass Chair in Physics of Complex Systems at Tel Aviv University, by the NSF-sponsored Center for Theoretical Biological Physics (CTBP), grants PHY-0216576 and 0225630 and by the University of California at San Diego.

## Appendix A

For the sake of simplicity we have adopted throughout the text the following notation for the generic Hill function:

$$\text{Hill}(x^n, K) \equiv \frac{x^n}{x^n + K^n}$$

where $n$ is the Hill coefficient and $K$ is the midpoint of the Hill function, namely the value of $x$ at which $\text{Hill}(x^n, K)\big|_{x=K} = 1/2$.

It can be shown that the product of two Hill functions can be approximated by the Hill function with the greatest midpoint, when the two midpoints are distant enough from each others, that is:

$$\text{Hill}(x^n, K_1) \cdot \text{Hill}(x^n, K_2) \approx \text{Hill}(x^n, K_2)$$

if and only if $K_1 << K_2$ (Figure 1, Online Supplementary Material). Indeed under such conditions $\text{Hill}(x^n, K_1) \cdot \text{Hill}(x^n, K_2) >> 0$ only when $x >> K_1$, hence

$$\text{Hill}(x^n, K_1) \cdot \text{Hill}(x^n, K_2) = \frac{x^{2n}}{x^{2n} + (K_1^n + K_2^n)x^n + K_1^n K_2^n} \approx \frac{x^{2n}}{x^{2n} + K_2^n x^n} = \text{Hill}(x^n, K_2)$$

This result can be extended to the product of $N$ Hill functions, that is:

$$\prod_{i=1}^{N} \text{Hill}(x^n, K_i) \approx \text{Hill}(x^n, \max(K_1, \ldots, K_N))$$

provided that $K_1 << K_2 << \ldots << K_N$.

Notably, the product of Hill function is not the only case in which a function composed by Hill functions can be approximated by a mere Hill function: other examples are given by functions of the type $\text{Hill}\big((\text{Hill}(x^n, K_1))^m, K_2\big)$ or $\text{Hill}\big(x^m, K_1 \cdot \text{Hill}(x^n, K_2)\big)$ (see Figure 2 in Online Supplementary Material).

## Appendix B

We seek an expression for [CaMKII*] based on the following kinetic reaction scheme:

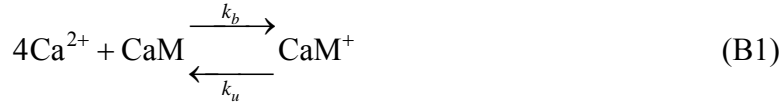

$$4\text{Ca}^{2+} + \text{CaM} \underset{k_u}{\overset{k_b}{\rightleftarrows}} \text{CaM}^+ \qquad (B1)$$

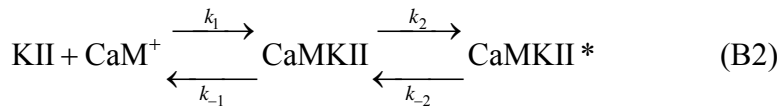

$$\text{KII} + \text{CaM}^+ \underset{k_{-1}}{\overset{k_1}{\rightleftarrows}} \text{CaMKII} \underset{k_{-2}}{\overset{k_2}{\rightleftarrows}} \text{CaMKII}^* \qquad (B2)$$

Let us first consider the reaction chain (B2). We can assume that the second step is very rapid with respect to the first one (De Konick and Schulman, 1998; Thiel et al., 1988) so that generation of CaMKII* is in equilibrium with CaMKII consumption, namely:

$$[\text{CaMKII}^*] \approx \frac{k_2}{k_{-2}}[\text{CaMKII}] \quad (B3)$$

Then, under the hypothesis of quasi-steady state for CaMKII, we can write:

$$\frac{d}{dt}[\text{CaMKII}] = k_1[\text{KII}][\text{CaM}^+] - (k_{-1} + k_2)[\text{CaMKII}] + k_{-2}[\text{CaMKII}^*] \approx 0 \quad (B4)$$

It follows that incorporation of equation (B3) into (B4) leads to:

$$[\text{CaMKII}^*] = K_1 K_2 [\text{KII}][\text{CaM}^+] \quad (B5)$$

where $K_i = k_i/k_{-i}$. Defining $[\text{KII}]_T = [\text{KII}] + [\text{CaMKII}] + [\text{CaMKII}^*]$ as the total kinase II concentration and assuming it constant, we can rewrite equation (B5) as follows:

$$[\text{CaMKII}^*] = \frac{K_2[\text{KII}]_T}{1+K_2} \frac{[\text{CaM}^+]}{[\text{CaM}^+] + K_m} \quad (B6)$$

with $K_m = (K_1(K_2 + 1))^{-1}$.

The substrate concentration for the enzymatic reaction (B2) is provided by reaction (B1) according to which:

$$[\text{CaM}^+] = [\text{CaM}] \frac{[\text{Ca}^{2+}]^4}{[\text{Ca}^{2+}]^4 + K_d} \quad (B7)$$

with $K_d = k_u/k_b$. Therefore, substituting equation (B7) into (B6), we obtain:

$$[\text{CaMKII}^*] = \frac{K_2[\text{KII}]_T}{1+K_2}\left(1 + \frac{K_m}{[\text{CaM}]}\right)^{-1} \frac{[\text{Ca}^{2+}]^4}{[\text{Ca}^{2+}]^4 + \dfrac{K_m K_d}{K_m + [\text{CaM}]}} \quad (B8)$$

so that $[\text{CaMKII}^*] \propto \text{Hill}([Ca^{2+}]^4, K_D)$ with $K_D = \left(\dfrac{K_m K_d}{K_m + [\text{CaM}]}\right)^{1/4}$.

**Table 1**

| Parameter | | Description | AM | FM |
|---|---|---|---|---|
| *L-R core*[1] | | | | |
| $r_C$ | s$^{-1}$ | Maximal CICR rate | 6 | |
| $r_L$ | s$^{-1}$ | Maximal rate of Ca$^{2+}$ leak from the ER | 0.11 | |
| $C_0$ | μM | Total cell free Ca$^{2+}$ concentration referred to the cytosol volume | 2 | |
| $c_1$ | - | Ratio between cytosol volume and ER volume | 0.185 | |
| $v_{ER}$ | μMs$^{-1}$ | Maximal rate of SERCA uptake | 0.9 | |
| $K_{ER}$ | μM | SERCA Ca$^{2+}$ affinity | 0.1 | 0.05 |
| $d_1$ | μM | IP$_3$ dissociation constant | 0.13 | |
| $d_2$ | μM | Ca$^{2+}$ inactivation dissociation constant | 1.049 | |
| $d_3$ | μM | IP$_3$ dissociation constant | 0.9434 | |
| $d_5$ | μM | Ca$^{2+}$ activation dissociation constant | 0.08234 | |
| $a_2$ | s$^{-1}$ | IP$_3$R binding rate for Ca$^{2+}$ inhibition | 0.2 | |
| *Agonist-independent IP$_3$ production*[2] | | | | |
| $\bar{v}_\delta$ | μMs$^{-1}$ | Maximal rate of IP$_3$ production by PLCδ | 0.02 | 0.05 |
| $K_{PLC\delta}$ | μM | Ca$^{2+}$ affinity of PLCδ | 0.1 | |
| $\kappa_\delta$ | μM | Inhibition constant of PLCδ activity | 1.5 | |
| *IP$_3$ degradation*[3] | | | | |
| $\bar{r}_{5P}$ | s$^{-1}$ | Maximal rate of degradation by IP-5P | 0.04 | 0.05 |
| $\bar{v}_{3K}$ | μMs$^{-1}$ | Maximal rate of degradation by IP$_3$-3K | 2 | |
| $K_D$ | μM | Ca$^{2+}$ affinity of IP$_3$-3K | 0.7 | |
| $K_3$ | μM | IP$_3$ affinity of IP$_3$-3K | 1 | |
| *Agonist-dependent IP$_3$ production*[4] | | | | |
| $\bar{v}_\beta$ | μMs$^{-1}$ | Maximal rate of IP$_3$ production by PLCβ | 0.2 | 0.5 |
| $K_R$ | μM | Glutamate affinity of the receptor | 1.3 | |
| $K_P$ | μM | Ca$^{2+}$/PKC-dependent inhibition factor | 10 | |
| $K_\pi$ | μM | Ca$^{2+}$ affinity of PKC | 0.6 | |

[1] De Pittà et al., (2008a); Li and Rinzel, (1994).

[2] Höfer et al., (2002); Rebecchi and Pentyala (2000); Pawelczyk and Matcki, (1997).

[3] De Konick and Schulman, (1998); Sims and Allbritton, (1998); Takazawa et al., (1989); Irvine et al., (1986).

[4] Suzuki et al., (2004); Höfer et al., (2002); Kawabata et al., (1996); Shinomura et al., (1991).

**Table 1**. Parameter values for the *ChI* and G-*ChI* models.

# FIGURE CAPTIONS

**Figure 1** (colors online). Block diagrams of *(a)* production and *(b)* degradation of inositol 1,4,5-trisphosphate (IP$_3$), summarize the complexity of the signaling network underlying glutamate-induced intracellular dynamics of this second messenger. A peculiar feature of IP$_3$ metabolism is its coupling with intracellular calcium (Ca$^{2+}$) dynamics which, in astrocytes, primarily occurs through *(c)* Ca$^{2+}$-induced Ca$^{2+}$ release (CICR) from intracellular stores. Production of IP$_3$ is brought forth by hydrolysis of the highly phosphorylated membrane lipid phosphatidylinositol 4,5-bisphosphate (PIP$_2$) by PLCβ and PLCδ, two isoenzymes of the family of phosphoinositide-specific phospholipase C. *(d)* PLCδ signaling is agonist-independent and modulated by Ca$^{2+}$. *(e)* The contribution of PLCβ to IP$_3$ production instead depends on agonist binding to G-protein coupled metabotropic receptors (mGluRs) found on the surface of the cell. Degradation of IP$_3$ mainly occurs through phosphorylation into inositol 1,3,4,5-tetrakisphosphate (IP$_4$), catalyzed by IP$_3$ 3-kinase (3K) and dephosphorylation by inositol polyphosphate 5-phosphatase (5P). The activity of IP$_3$-3K is regulated by Ca$^{2+}$ in a complex fashion which may be approximated as depicted in *(f)*. For simplicity, inhibition of IP-5P by Ca$^{2+}$/CaMKII-dependent phosphorylation (Communi et al., 2001) and competitive binding of IP$_4$ to IP-5P are not considered in this study. Legend of the different arrows is below *(f)*.

**Figure 2** (colors online). Both AM-encoding or FM-encoding Ca$^{2+}$ oscillations can be brought forth by the Li-Rinzel (L-R) model for CICR, depending on the value of $K_{ER}$, the Ca$^{2+}$ affinity of (Sarco-)Endoplasmic-Reticulum Ca$^{2+}$-ATPase (SERCA) pumps. For example AM-encoding can be found at *(a-c)* $K_{ER}$=0.1 μM whereas FM-encoding exists for smaller $K_{ER}$, such as *(d-f)* $K_{ER}$=0.05 μM. *(a)* In the phase plane, AM-encoding is associated with a single intersection between *C*-nullcline (*orange*) and *h*-nullcline (*green*) (i.e. the curves for which $\dot{C} = 0$ and $\dot{h} = 0$ respectively). Accordingly, the only possible bifurcations that can be found are connected with loss/gain of stability, namely they are Hopf bifurcations. *(b)* The associated bifurcation diagram indeed shows that oscillations rise via supercritical Hopf bifurcation (H$_1$) at [Ca$^{2+}$]≈0.15 μM and [IP$_3$] ≈0.36 μM, whereas they die at $[Ca^{2+}]\approx 0.32$ μM and $[IP_3] \approx 0.64$ μM via subcritical Hopf bifurcation ($H_2$). The fact that $H_1$ is supercritical accounts for the occurrence of oscillations of arbitrarily small amplitude that increases with $IP_3$ yet *(c)* with almost constant period. *(d)* In FM-encoding conditions instead, the *C*-nullcline is sharply N-shaped and there exists a small range of $IP_3$ values where it can intersect the *h*-nullcline at three points. *(e)* This region is delimited by two "knees" showed by the fixed-point continuation curve, which correspond to a saddle-node bifurcation at $[Ca^{2+}]\approx 0.13$ μM and $[IP_3] \approx 0.48$ μM and a saddle-node homoclinic bifurcation (SNHom) at $[Ca^{2+}]\approx 0.07$ μM and $[IP_3] \approx 0.53$ μM. Pulsatile oscillations rise and die via subcritical Hopf bifurcations respectively at $H_1$ ($[Ca^{2+}]\approx 0.05$ μM, $[IP_3] \approx 0.51$ μM) and $H_2$ ($[Ca^{2+}]\approx 0.39$ μM, $[IP_3] \approx 0.86$ μM). While their amplitude is essentially constant, *(f)* their period can be arbitrarily long due to the homoclinic of the SNHom. *(b,e)* Conventions: stable equilibria are shown as *full lines*, respectively for low (*black*) and high $IP_3$ (*blue*) concentrations. Unstable equilibria are displayed as *red dashed lines*. Oscillations are located in the diagram as min (*green*)–max (*black*) envelopes, with stable oscillations as *full circles* and unstable ones as *empty circles*. Parameter values for the L-R model as in Table 1.

**Figure 3** (colors online). Bifurcation diagrams for PLCδ-dependent $IP_3$ production are drawn by substituting into equation (7), $[Ca^{2+}]$ and $[IP_3]$ values obtained from the bifurcation diagrams *(a)* in Figure 2b and *(b)* in Figure 2e, respectively. Colors as in Figures 2b,e.

**Figure 4** (colors online). *(a)* Experimental observations hint the existence of three regimes of $IP_3$ metabolism: one for low $[Ca^{2+}]$ and $[IP_3]$ in which $IP_3$-3K ($Ca^{2+}$-dependent *color* curves) and IP-5P (*black* curve) activities are similar; an intermediate one for higher $[Ca^{2+}]$ in which $IP_3$ degradation by $IP_3$-3K is predominant; and a third one for $[IP_3]$>8 μM in which $IP_3$ is degraded mainly by IP-5P in a $Ca^{2+}$-independent fashion. Both enzymes can be assumed Michaelis-Mentenian. *(b-c)* Physiological $IP_3$ concentrations suggest to consider only the first two regimes. Notably, these latter can be mimicked either *(b)* by keeping the hypothesis of Michaelis-Menten kinetics for $IP_3$-3K (equation 9) or *(c)* by a linear approximation of this dependence (equation 13).

**Figure 5** (colors online). Bifurcation behaviors of IP$_3$-3K-dependent IP$_3$ degradation in *(a-b)* AM and *(c-d)* FM conditions are compared for *(a,c)* Michaelis-Menten, (equation 12) or *(b,d)* linear approximations (equation 13) of the IP$_3$ dependence of IP$_3$-3K rate. Despite qualitatively similar behaviors, the linear approximation is not further taken into account in the present study, because IP$_3$-3K activity may saturate in physiological conditions, thus invalidating the linear approximation.

**Figure 6** (colors online). Consistency of the equation for the endogenous IP$_3$ metabolism with respect to the L-R core model can be tested as follows. At resting physiological conditions: $\dot{C} = \dot{I} = 0$, $h = h_\infty(C)$ and $C \to 0$ so that $v_{3K}(C, I_s) \approx 0$. Hence, for steady IP$_3$ values ($I_s$) such as $I_s \ll \kappa_\delta$, one gets $v_\delta(C, I_s) \approx \bar{v}_\delta \cdot \text{Hill}(C^2, K_{PLC\delta})$. Accordingly, equation (14) can be solved for $I_s$ yielding $I_s(C) \approx r_{5P}^{-1} \cdot \bar{v}_\delta \cdot \text{Hill}(C^2, K_{PLC\delta})$ (*magenta* curve). The latter curve must be compared with the corresponding *I(C)* curve (*black*) obtained by solving for *I* the equation $\dot{C}\big|_{h=h_\infty(C)} = 0$ in the original L-R model (equation 5). By changing $\bar{v}_\delta$, $\bar{r}_{5P}$, $K_{PLC\delta}$ and $\kappa_\delta$ according with to their experimental values, we seek consistency when $I_s(C) \approx I(C)$. In these conditions in fact our mathematical description of IP$_3$ metabolism and the L-R model predict equivalent steady intracellular IP$_3$ levels.

**Figure 7** (colors online). Surfaces for $\dot{C} = 0$ (*orange*), $\dot{h} = 0$ (*green*) and $\dot{I} = 0$ (*red*) for the *ChI* model described by equations (5), (6) and (14).

**Figure 8** (colors online). Projections of the surfaces for $\dot{C} = 0$ (*orange*), $\dot{h} = 0$ (*green*) and $\dot{I} = 0$ (*red*) onto the *I-C* plane for different values of *(a,c) h* or *(b,d) C*, both *(a,b)* in AM and *(c,d)* FM conditions, allow to appreciate the nature of coupling between IP$_3$ metabolism and Ca$^{2+}$ dynamics in the *ChI* model. In particular, since none of the parameters of the equation for IP$_3$ metabolism (equation 14) are to found in the equations for $\dot{C} = 0$ and $\dot{h} = 0$, the latter surfaces are not affected by inclusion of IP$_3$ dynamics into the L-R core model. It follows that IP$_3$ dynamics may influence Ca$^{2+}$ dynamics only through modulations of the dynamics of *h*, i.e. Ca$^{2+}$-mediated deactivation of CICR IP$_3$R/channels.

**Figure 9** (colors online). *(a-b)* Numerical investigation shows that the term related to the GTPase-dependent PLCβ termination pathway in the expression of the agonist-dependent IP$_3$ production (equations 15 and 17) can be neglected so that $K_\gamma(\gamma,C) \approx K_\gamma(C)$. *(c-d)* Bifurcation behaviors of $v_{glu}(\gamma,C)$ (equation 19), obtained by substituting $\gamma$ and *C* with their values derived from bifurcation diagrams of the agonist-dependent model (see also Figures 10a and 11a).

**Figure 10** (colors online). Bifurcations diagrams for AM-derived parameter sets of the G-*ChI* model (equations 5-6, 17), show *(c,f)* that the inclusion of IP$_3$ dynamics remarkably affects the frequency of oscillations. *(a,d)* In particular, Ca$^{2+}$ oscillations are essentially AFM-encoding rather than merely AM-encoding. *(d-f)* Low values of the glutamate-dependent maximal rate of IP$_3$ production $\bar{v}_\beta$, extend the range of oscillations to arbitrarily high glutamate concentrations. In these conditions phase-locked Ca$^{2+}$/IP$_3$ oscillations and pulsations can be observed. Namely, there is a threshold glutamate-concentration (which can equivalently be described by a threshold frequency of a pulsed stimulation), for which the frequency of oscillations (pulsations) locks to a particular value and does not change for further elevations of glutamate concentration. Parameters as in Table 1 except for *(d-f)* where $\bar{v}_\beta$=0.05 μMs$^{-1}$.

**Figure 11** (colors online). Bifurcation diagrams of the G-*ChI* model for FM-encoding sets of parameters. *(d-i)* In analogy with Figure 10, reduced values of $\bar{v}_\beta$, the maximal rate of PLCβ-dependent IP$_3$ production, extends to infinity the range of oscillations, leading to phase-locking of Ca$^{2+}$/IP$_3$ pulsating oscillations. *(d-f)* There is also an intermediate range of $\bar{v}_\beta$ values for which oscillations and fixed concentrations of [Ca$^{2+}$] and [IP$_3$] can coexist. *(b-c,e-f,h-i)* Unlike Ca$^{2+}$ oscillations, IP$_3$ oscillations are always

AFM encoding with respect to the concentration of agonist (see also Figures 10b-c,e-f). Parameters as in Table 1 except for *(d-f)*: $\bar{v}_\beta$=0.2 µMs$^{-1}$; and *(g-i)*: $\bar{v}_\beta$=0.05 µMs$^{-1}$.

**Figure 12**. *(a-b)* Examples of forced burst oscillations exhibited by the G-*ChI* model, under *(c)* a square-wave stimulus protocol. This figure illustrates how stationary glutamate stimulations are encoded as oscillations and pulsations of the second messengers Ca$^{2+}$ and IP$_3$. A closer look at oscillatory patterns in *(a-b)* reveals that in our model, IP$_3$ oscillations always lag Ca$^{2+}$ oscillations. Indeed, the adoption of the L-R core model for CICR at constant IP$_3$ concentration implies that IP$_3$ oscillations are not a prerequisite for Ca$^{2+}$ oscillations to occur. Square-wave stimulus: (AM): $\gamma_{min}$=2 nM, $\gamma_{max}$=5 µM; (FM): $\gamma_{min}$=1 nM, $\gamma_{max}$=6 µM; (AM, FM): duty cycle: 0.5. Note that in the FM case, the value of $\gamma_{max}$ corresponds in the bifurcation diagrams in Figures 11a-b to a bistable state (a stable fixed point and a stable limit cycle separated by an unstable limit cycle). This explains why pulsations at high stimulations are of limited duration.

**Figure 13**. Simulated Ca$^{2+}$ and IP$_3$ patterns obtained when the G-*ChI* model is fed with physiologically realistic glutamate stimulations, in the AM and FM case. A striking feature is a remarkable increase of the signal smoothness, when one goes from glutamate-stimulus to IP$_3$ traces *(b-c)* and from the latter to Ca$^{2+}$ traces *(a-b)*. This fact hints different integrative properties for IP$_3$ and Ca$^{2+}$ with respect to the stimulus, which are likely to be cross-coupled (see Discussion).

# ONLINE SUPPLEMENTARY MATERIAL

**Figure 1**. The product of two Hill functions (*a-b*) with sufficiently distant midpoints is equivalent to the Hill function with the largest midpoint (*c*). Namely: $\text{Hill}(x, K_1) \cdot \text{Hill}(x, K_2) \approx \text{Hill}(x, K_2)$ where $K_1 \ll K_2$. Midpoints are marked by vertical dashed lines; $K_1$: *red*; $K_2$: *blue*.

**Figure 2**. Hill functions of Hill functions (*a-b*) can also be approximated by Hill functions. (*c-d*) $\text{Hill}(\text{Hill}(x, K_2), K_1) = (1 + K_1)^{-1} \cdot \text{Hill}(x, K_1 K_2 (1 + K_1)^{-1})$. In this case the midpoint of the resulting Hill function depends on the specific values of the midpoints of the original Hill functions considered in the composition of the Hill-of-Hill function. (*e-h*) $\text{Hill}(x, K_1 \cdot \text{Hill}(x, K_2)) = (x + K_2)/(x + K_1 + K_2) = \text{Hill}(x, (K_1 + K_2)) + f(x)$, where $f(x) = K_2/(x + K_1 + K_2)$. Notably, $f(x \to 0) = K_2/(K_1 + K_2)$ whereas $f(x \to \infty) \approx 0$, so that the resulting Hill curve is essentially comprised within the interval $[K_2/(K_1 + K_2), 1)$.

**Figure 3**. Bifurcation diagrams for a modified *ChI* model and prototypical sets of *(a-c)* AM-encoding and *(d-f)* FM-encoding L-R parameters. The bifurcation diagrams were computed after introduction into the *ChI* model of the rate of glutamate-dependent IP$_3$ production, $v_{glu}$, as a free bifurcation parameter, namely $\dot{I} = v_{glu} + v_\delta(C, I) - v_{3K}(C, I) - v_{5P}(I)$. This figure shows that the *ChI* model can still display oscillations in presence of an external non-specific bias of IP$_3$ production. This is a first suggestion that the corresponding glutamate-dependent G-*ChI* model may also display oscillations. The parameters are taken from Table 1.

Figure 1

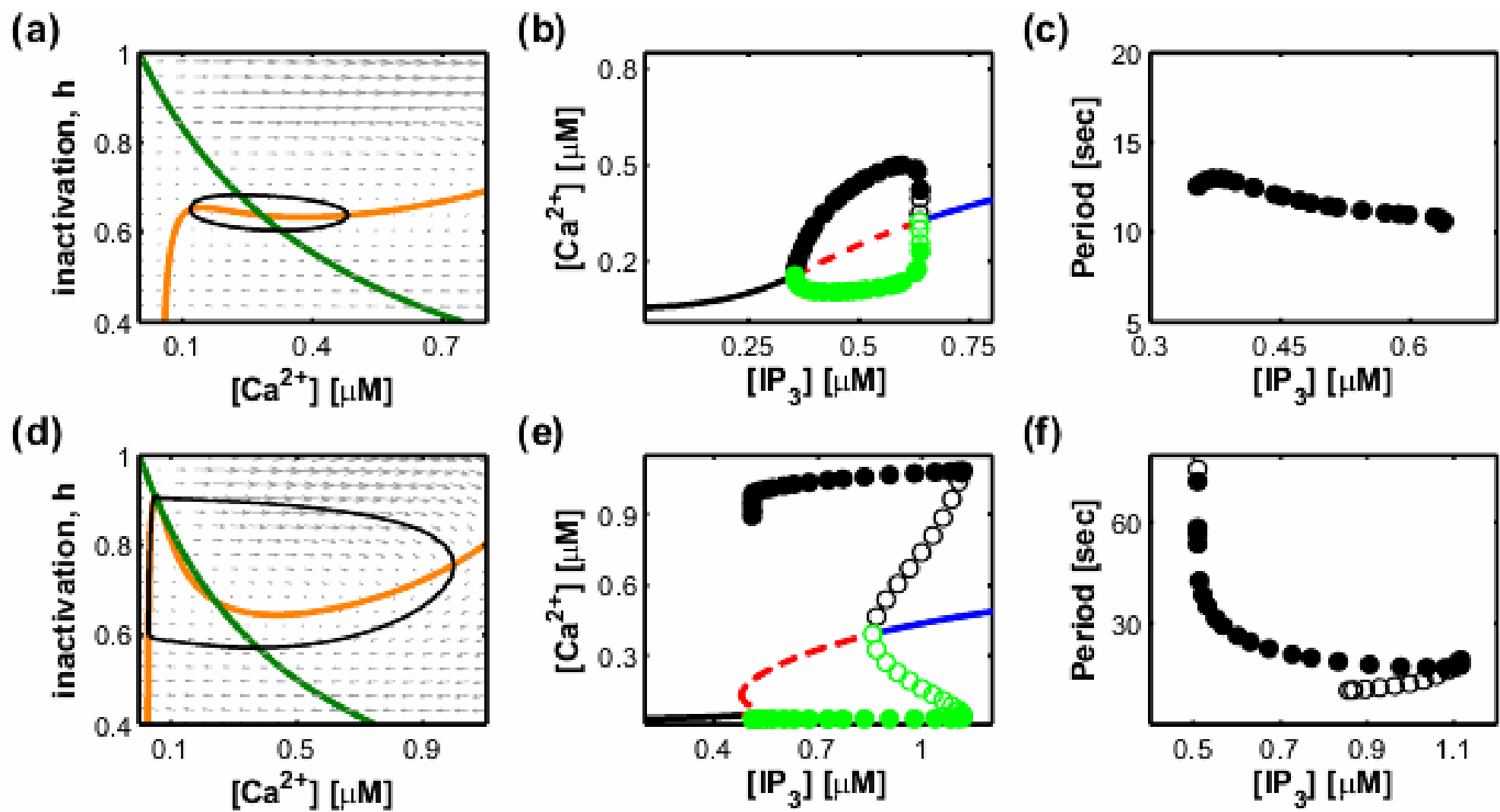

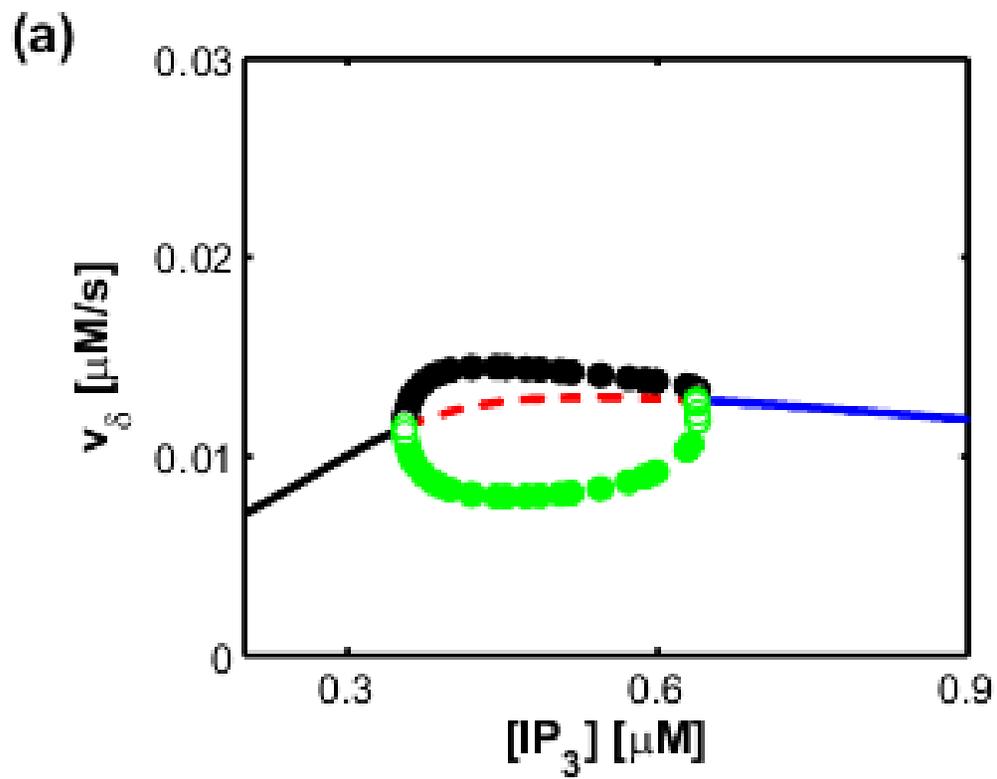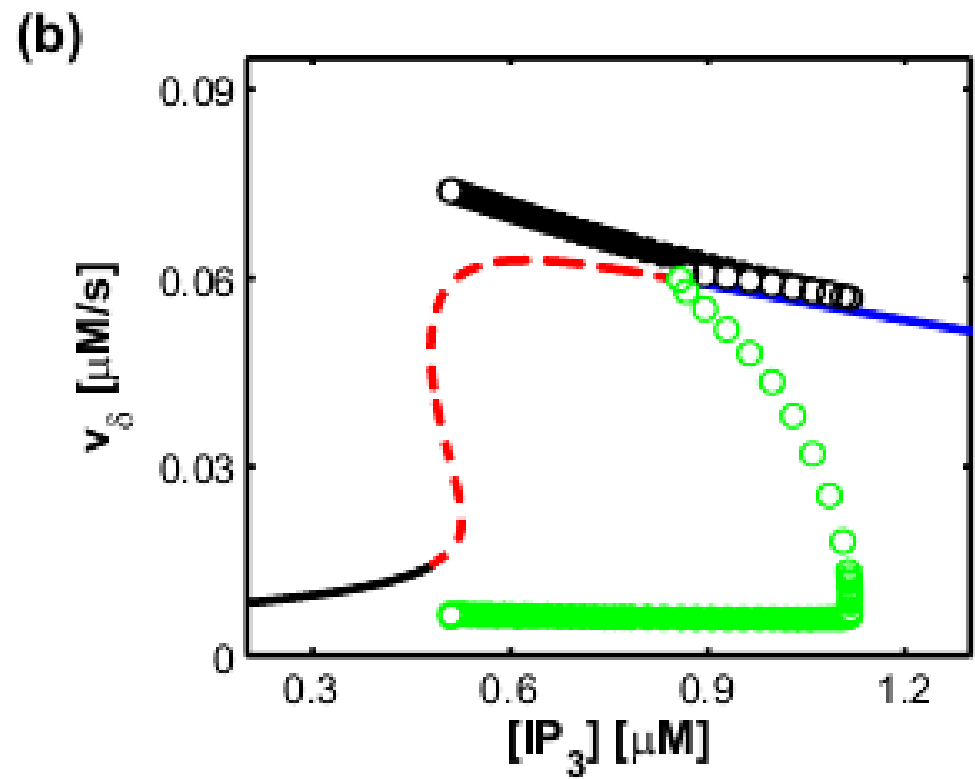

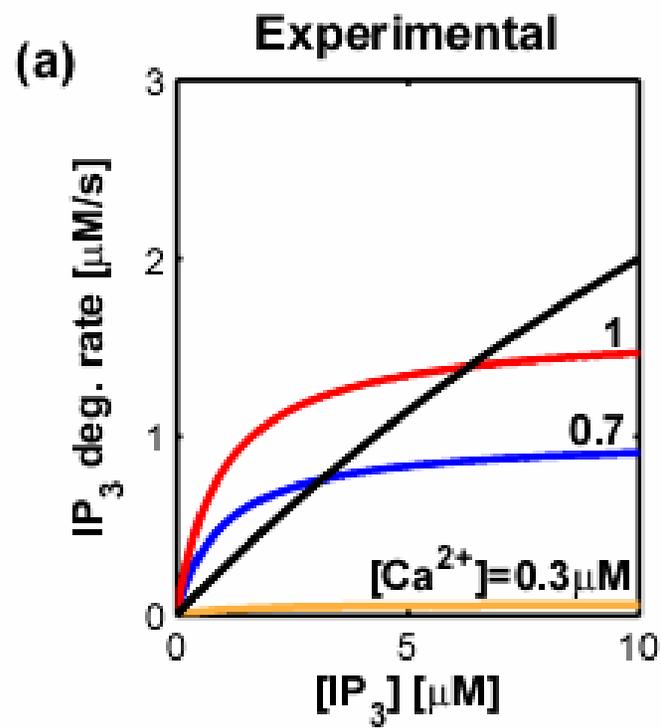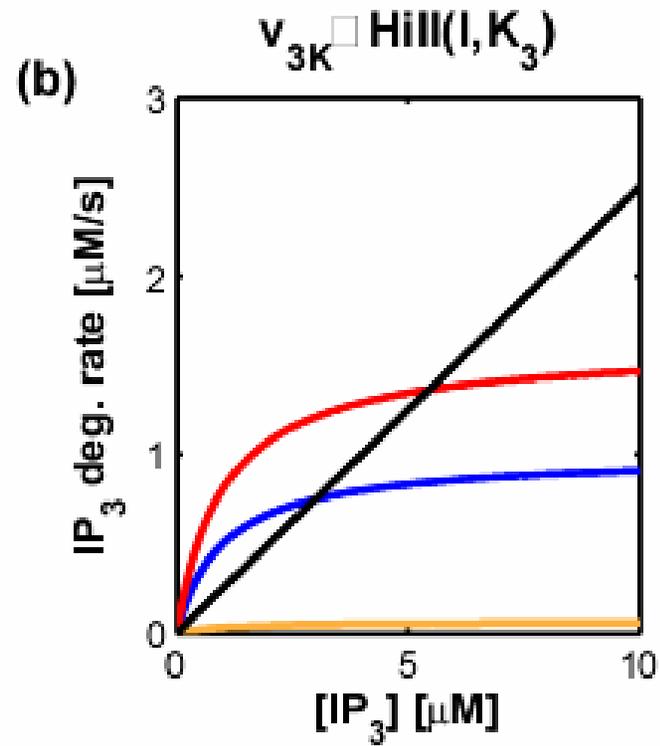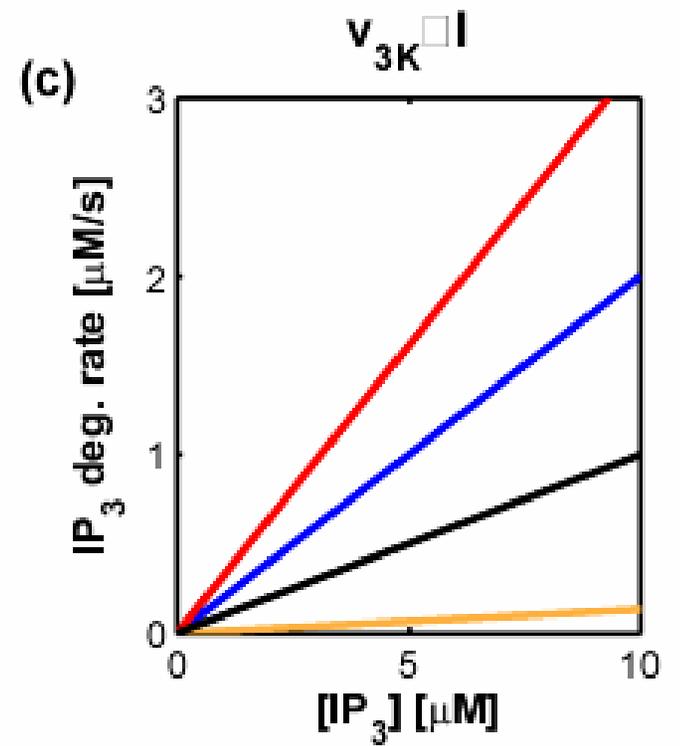

Figure 4

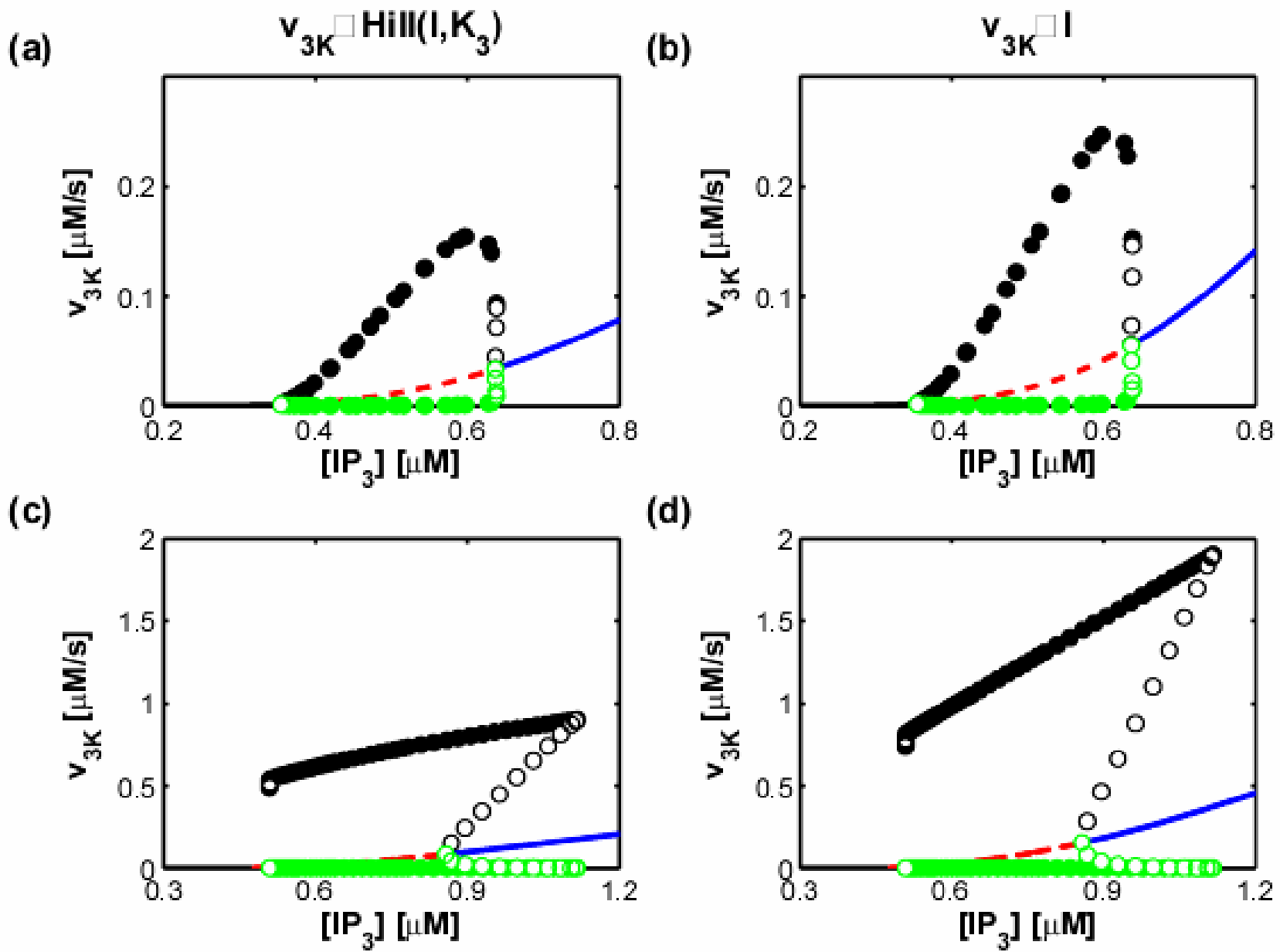

Figure 5

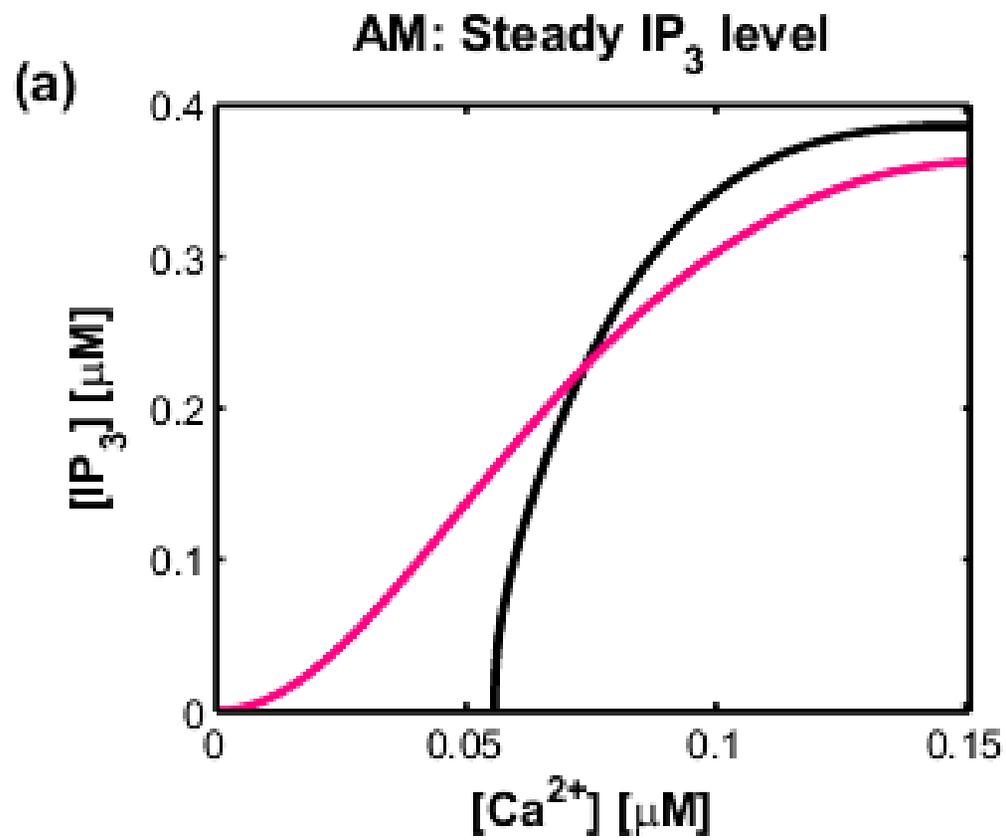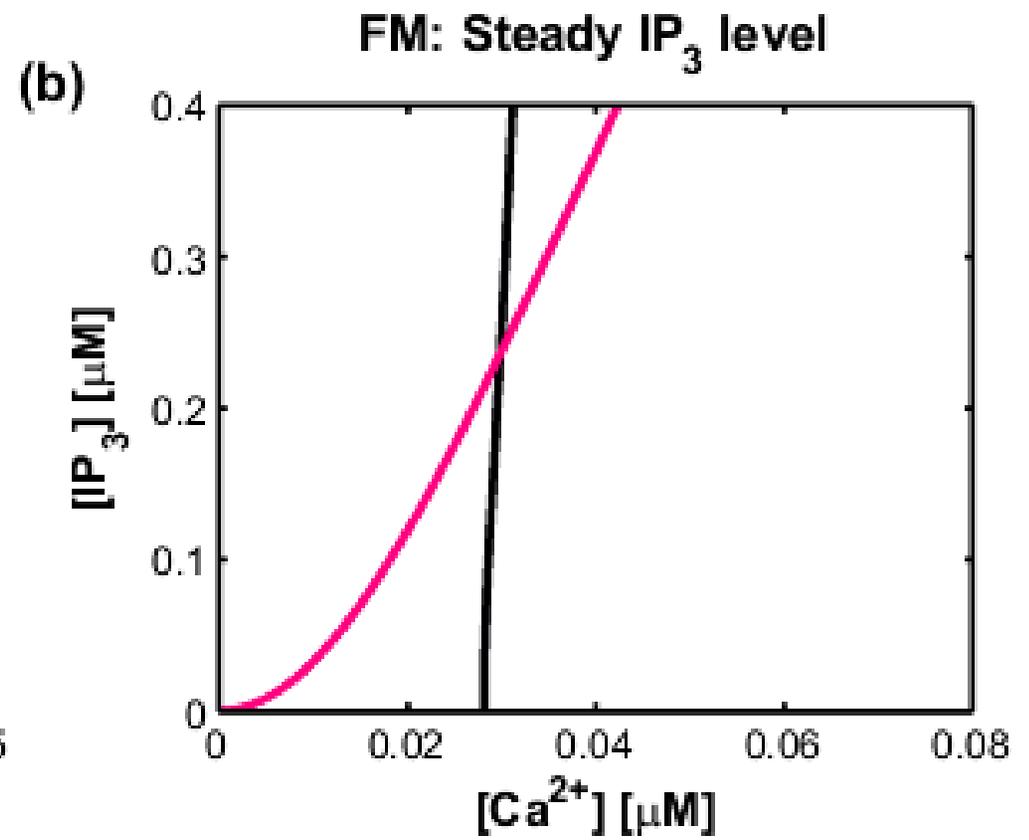

Figure 6

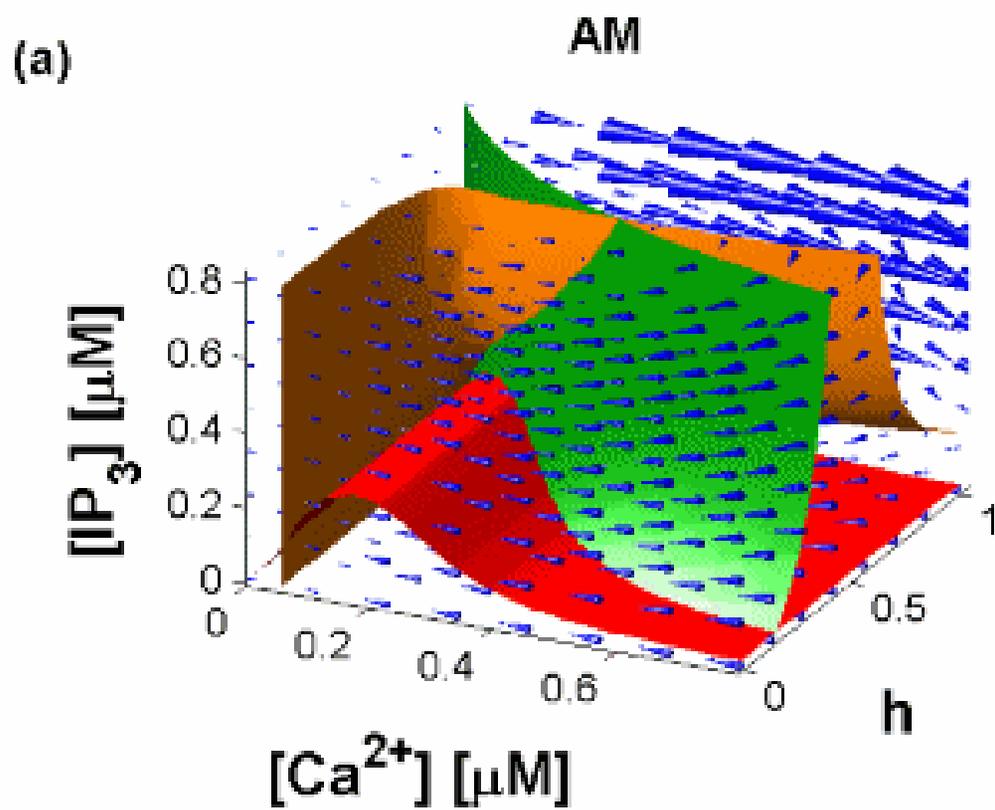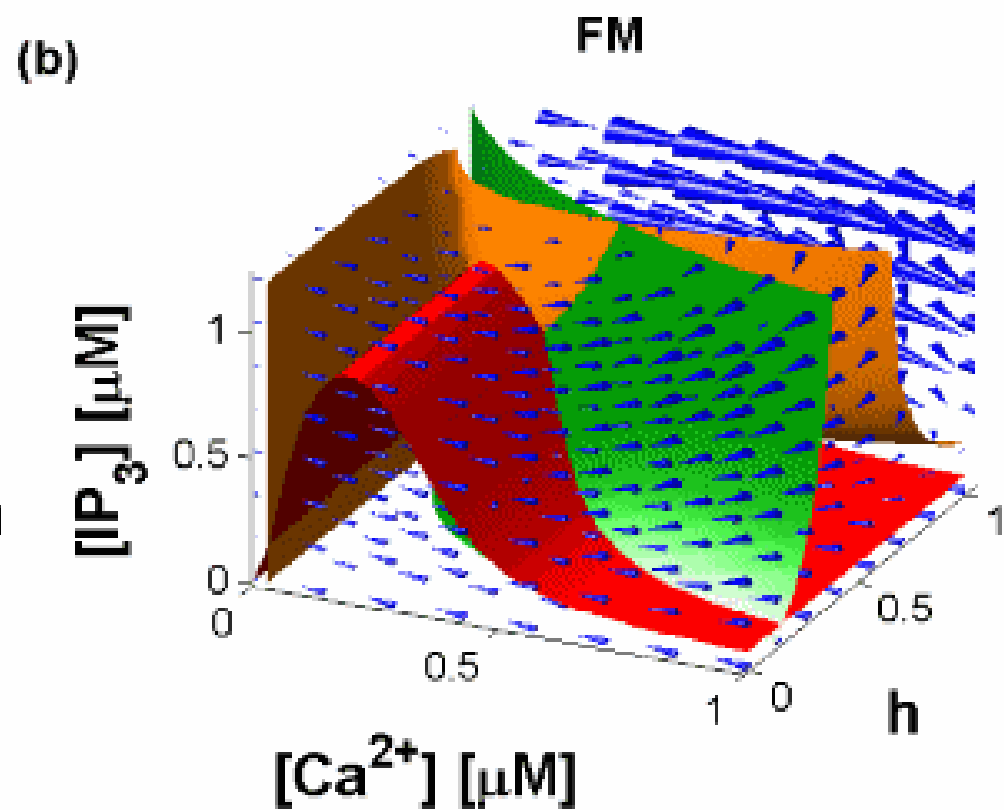

Figure 7

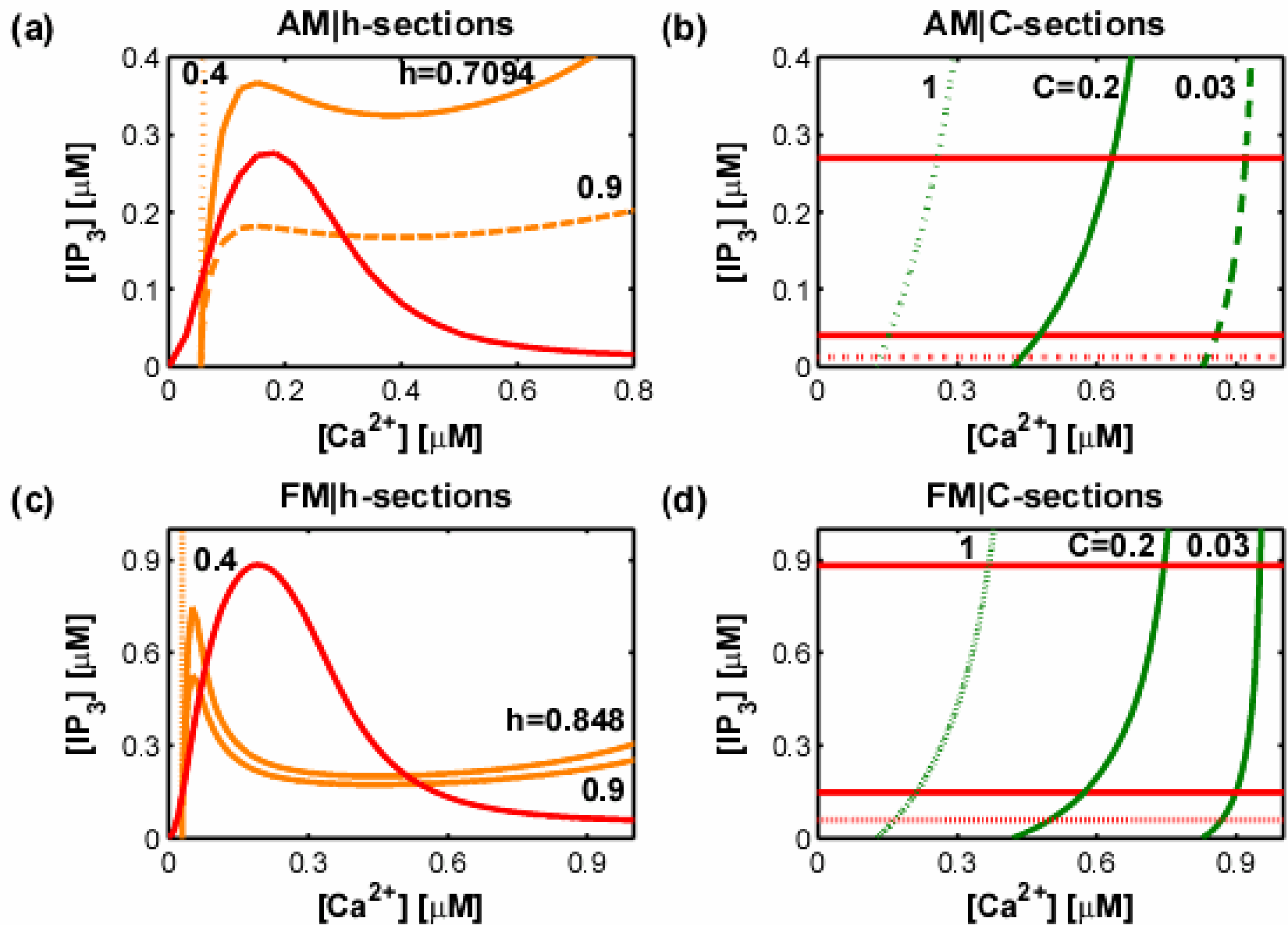

Figure 8

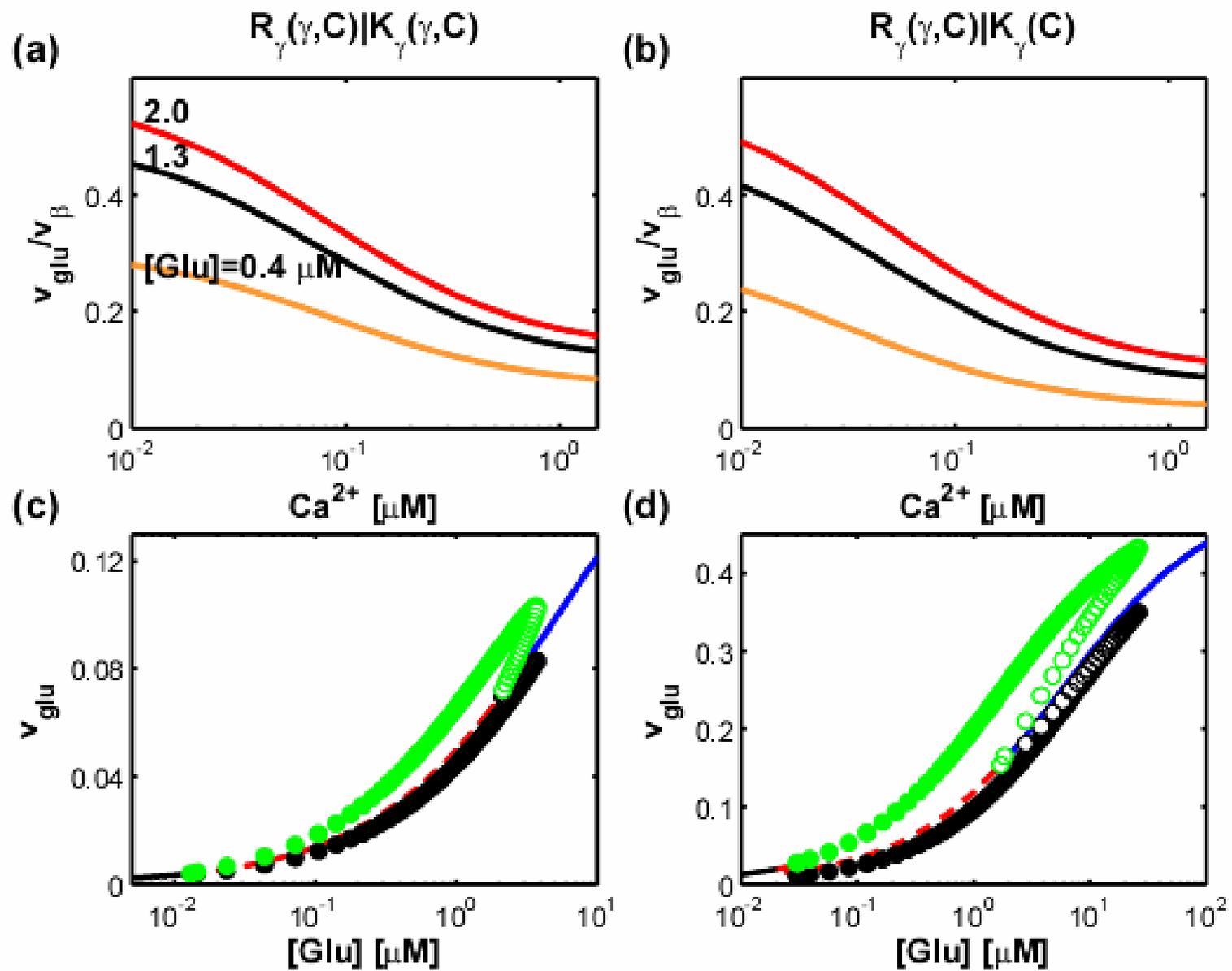

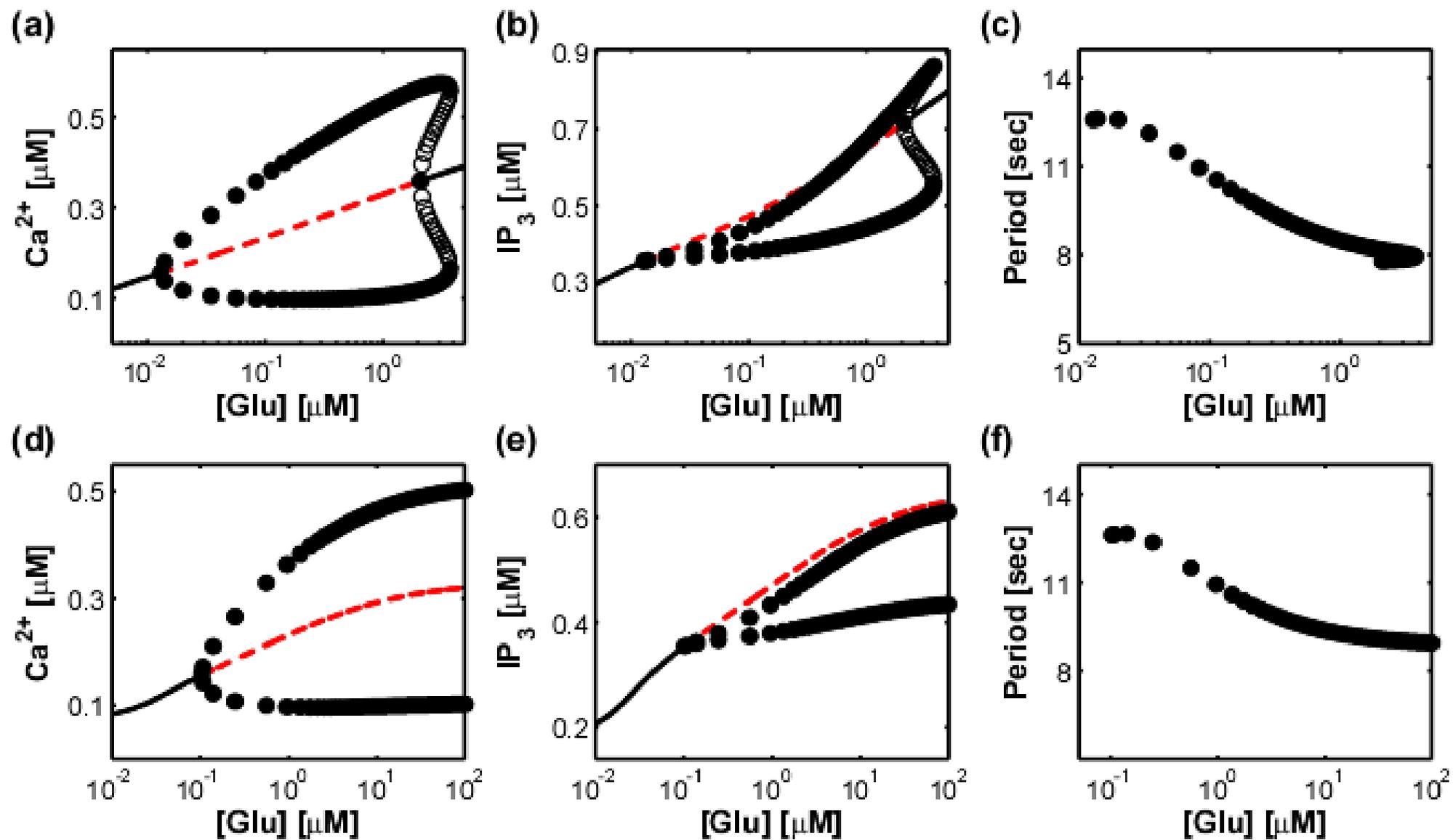

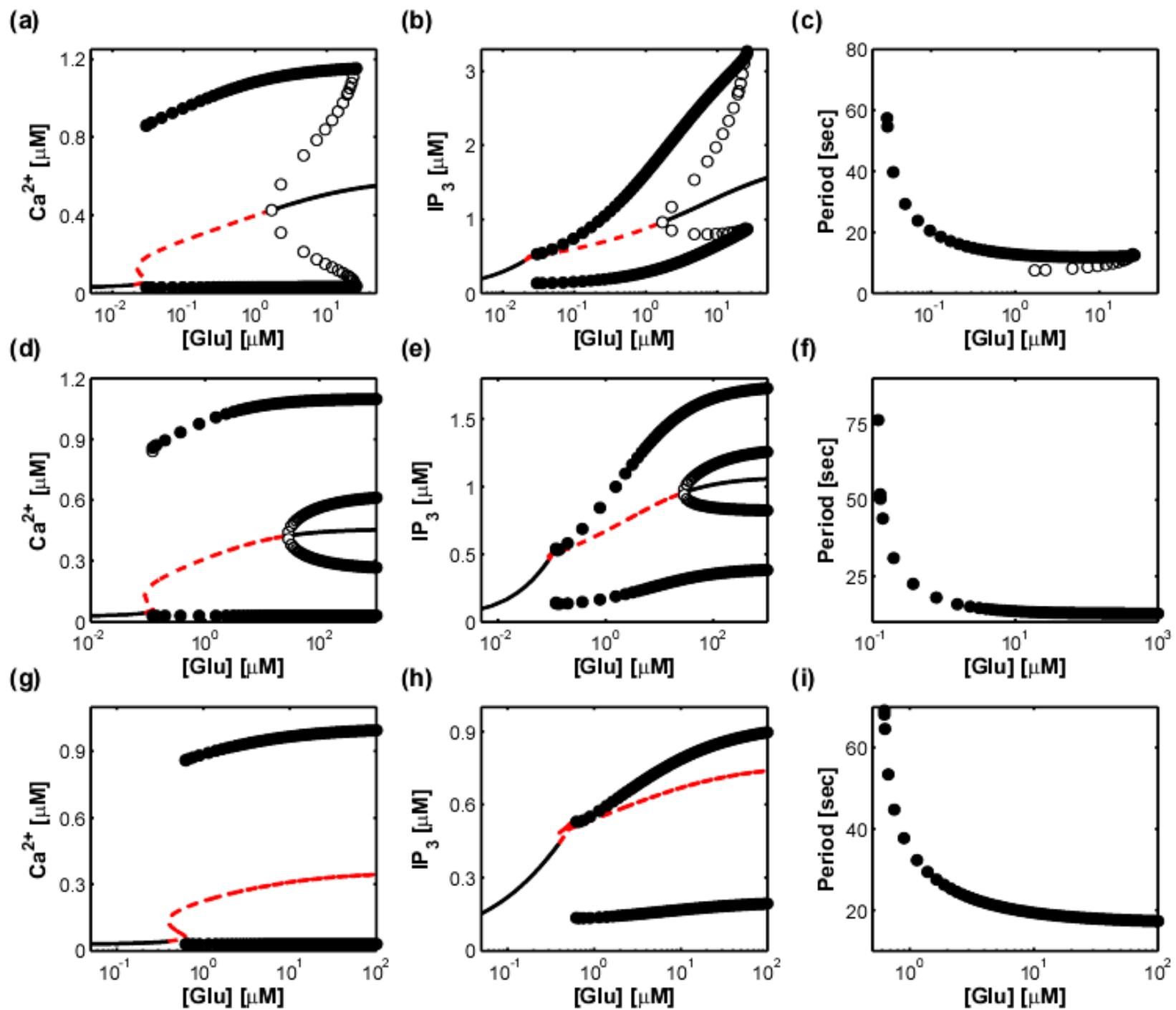

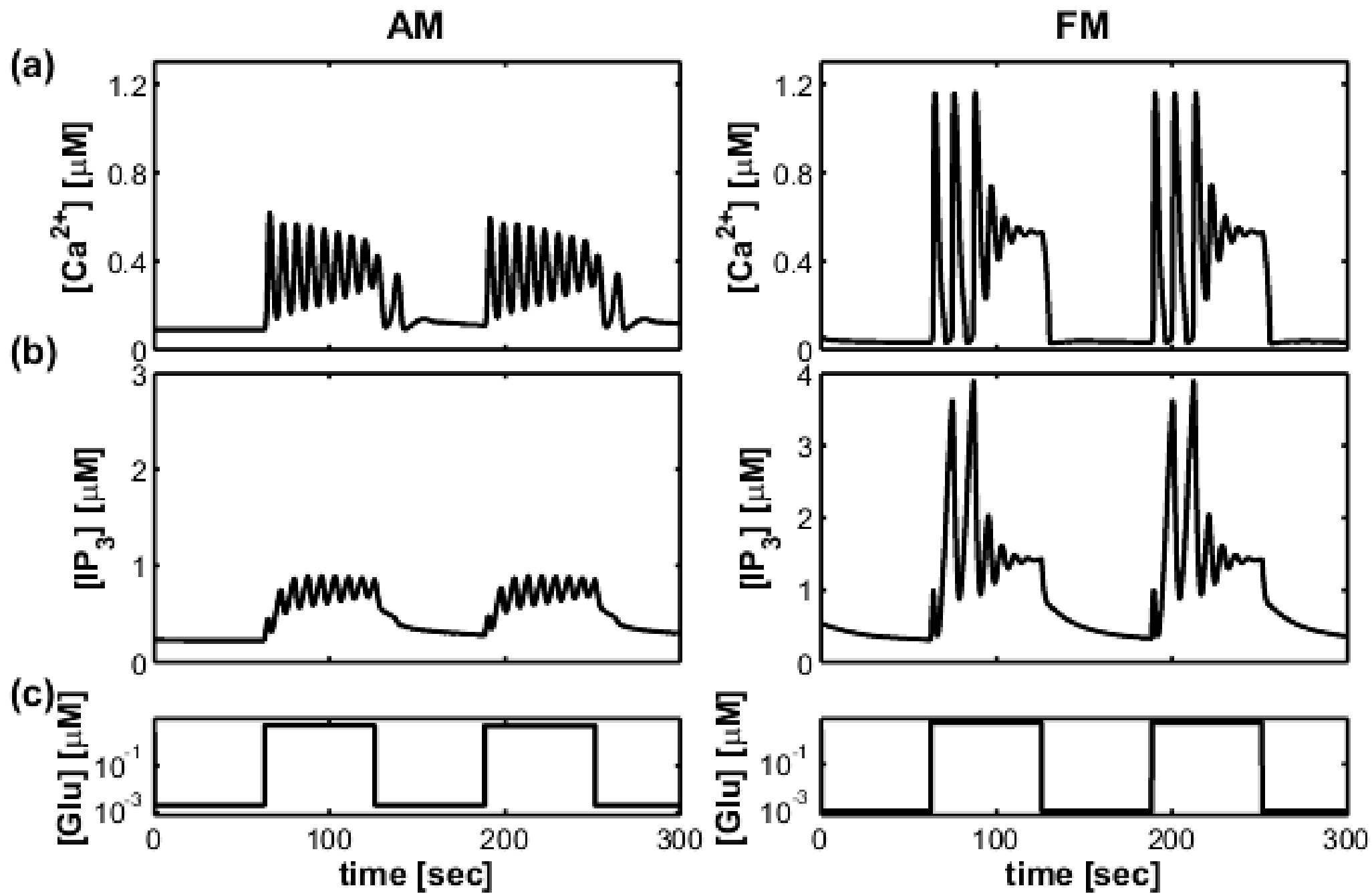

Figure 12

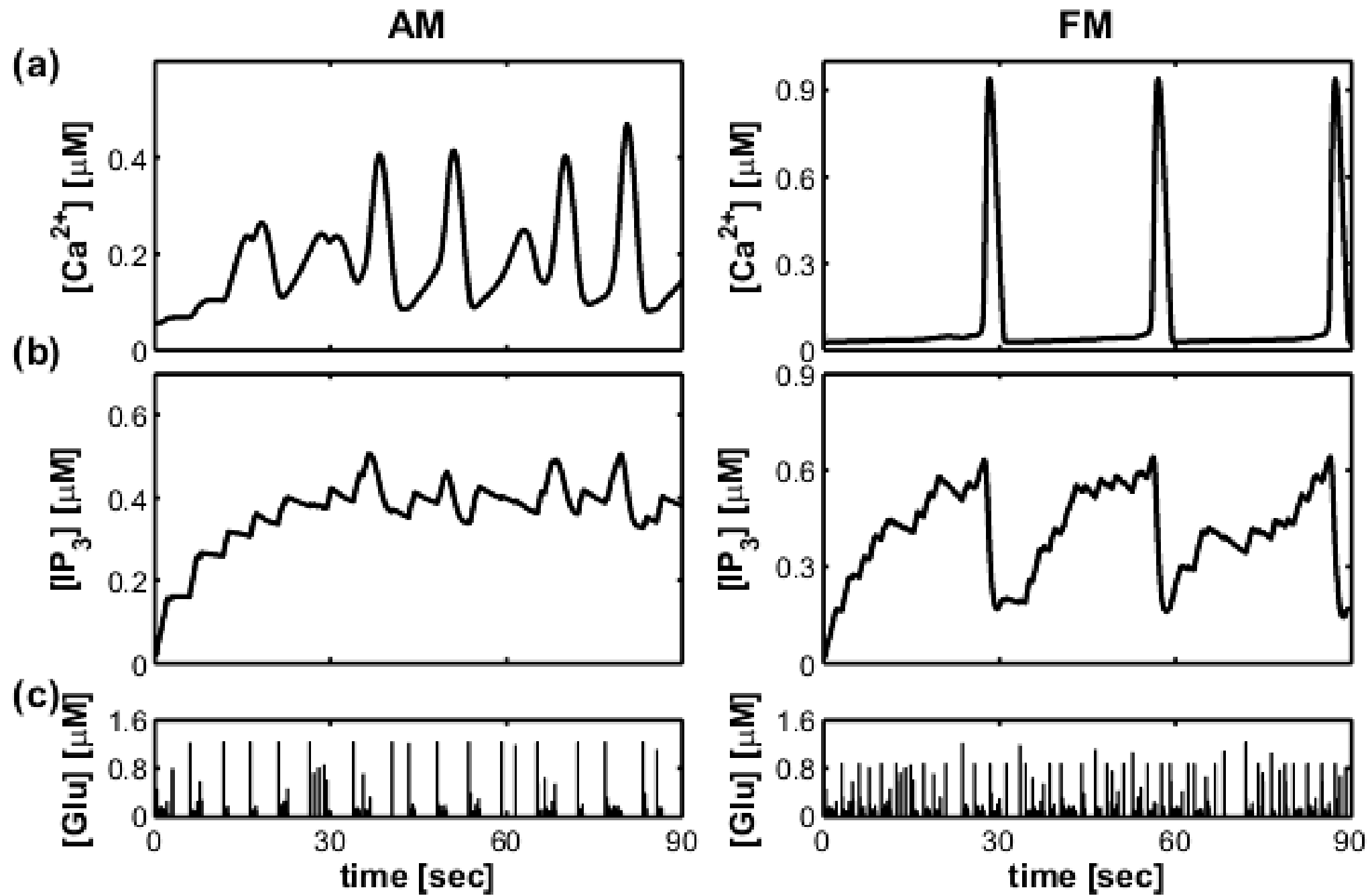

Figure 13

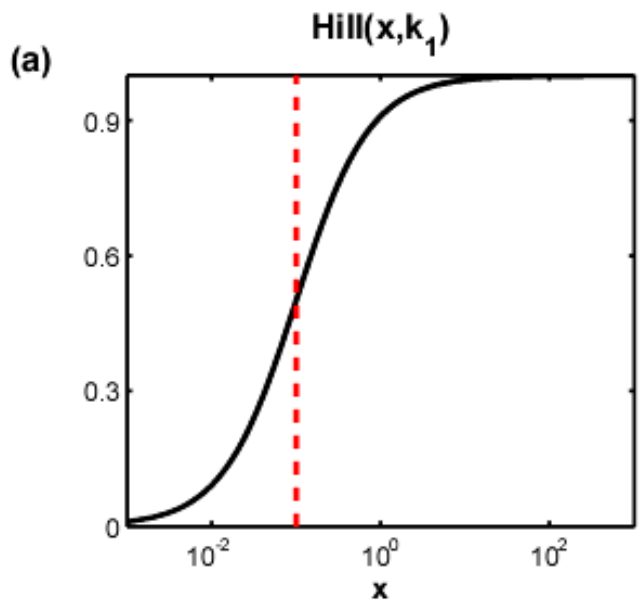 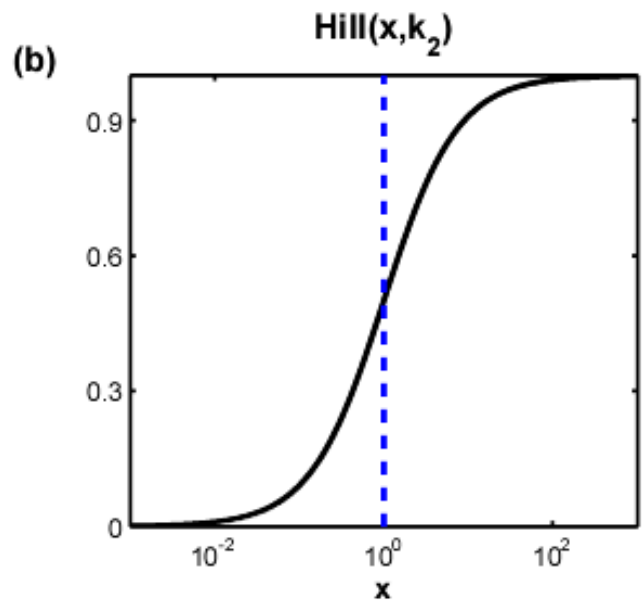 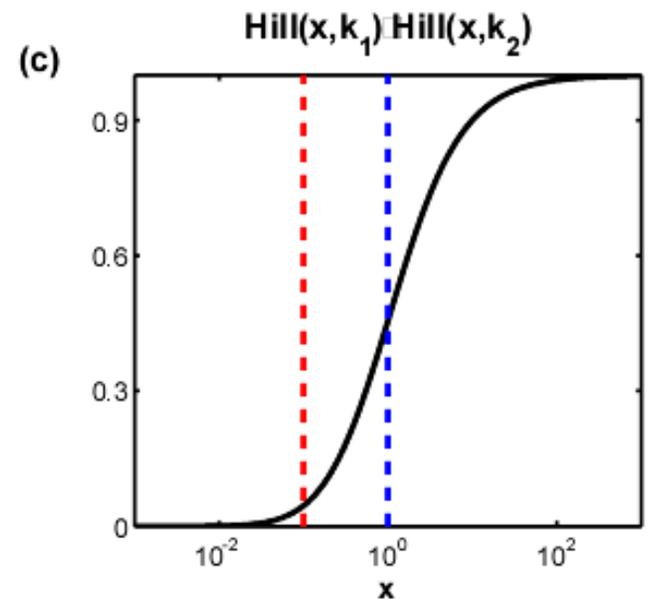

Supplementary Figure 14

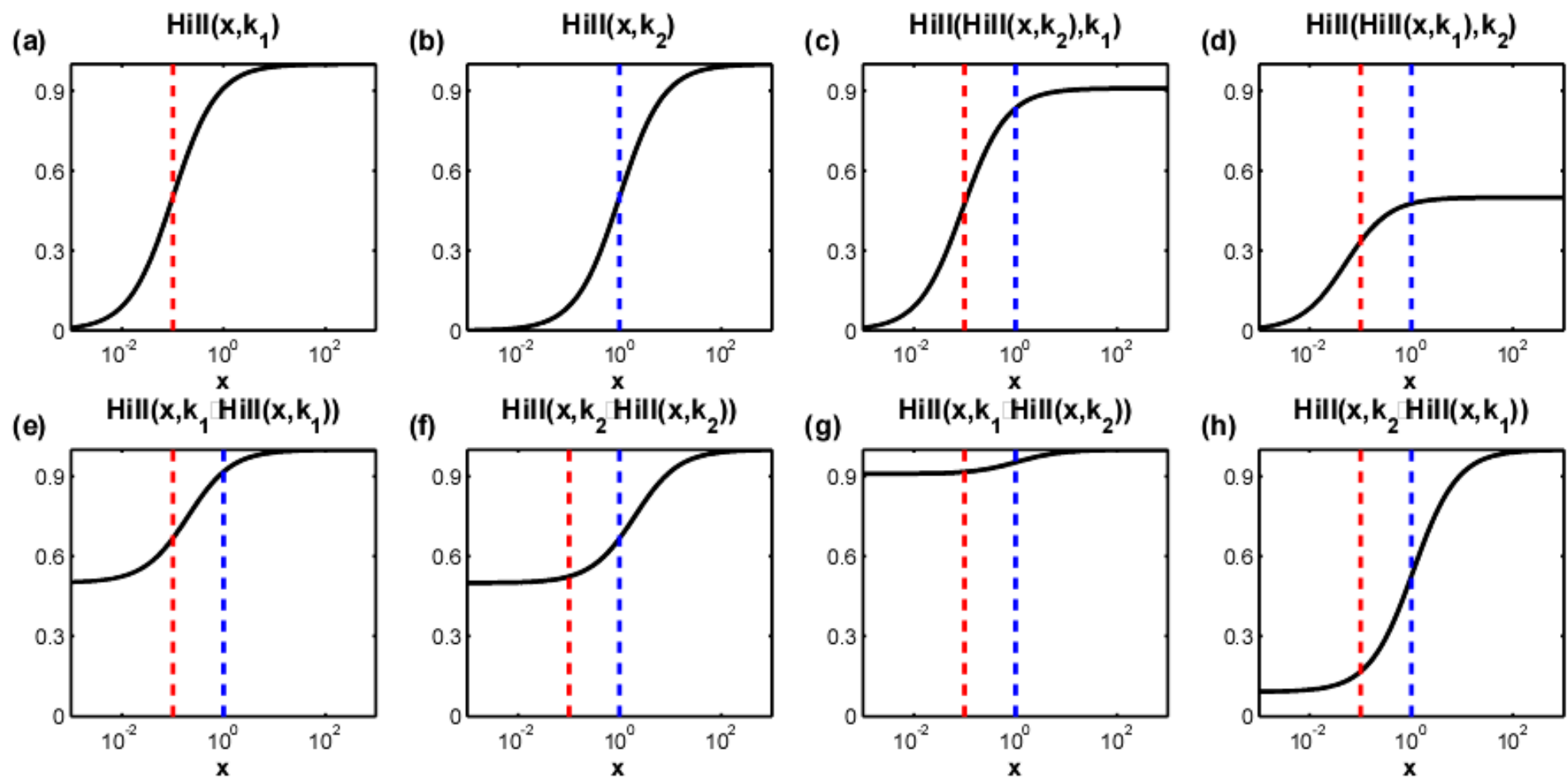

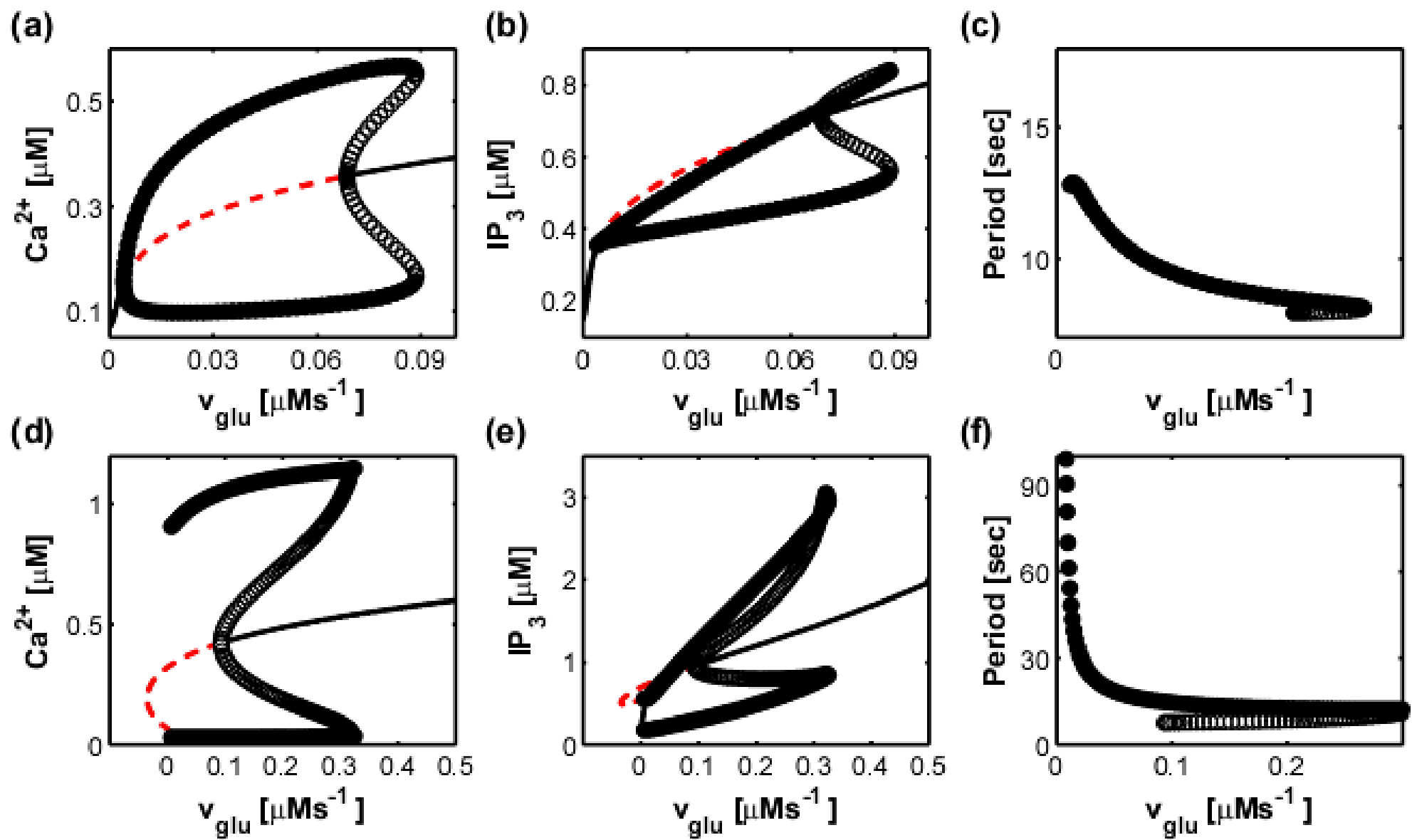